\documentclass[
superscriptaddress,
amsmath,
amssymb,
aps,
prb,
twocolumn,
nofootinbib
]{revtex4-2}

\usepackage{graphicx}         
\usepackage{dcolumn}          
\usepackage{bm}               
\usepackage{upgreek}
\usepackage{tabularx}
\usepackage[colorlinks=true,urlcolor=blue,breaklinks=true]{hyperref}

\usepackage{xcolor}

\usepackage{xr}
\newcommand{\lef}{\left(}
\newcommand{\rig}{\right)}

\newcommand{\Tr}{\mathrm{Tr}}
\newcommand{\deriv}{\mathrm{d}}
\newcommand{\sign}{\mathrm{sgn}}

\newcommand*{\rttensor}[1]{\bar{\bar{#1}}}
\newcommand{\Li}{\mathrm{Li}}

\newcommand{\e}{\mathrm e}
\newcommand{\kB}{k_\mathrm{B}}

\newcommand{\Le}{{\mathrm{L}}}
\newcommand{\Ri}{{\mathrm{R}}}

\newcommand{\vecB}{\mathbf B}
\newcommand{\vecE}{\mathbf E}
\newcommand{\vecj}{\mathbf j}
\newcommand{\vecn}{\mathbf n}
\newcommand{\vecp}{\mathbf p}
\newcommand{\vecS}{\mathbf S}
\newcommand{\vecr}{\mathbf r}
\newcommand{\vecu}{\mathbf u}
\newcommand{\vecV}{\mathbf V}
\newcommand{\vecalpha}{\bm{\alpha}}
\newcommand{\vecsigma}{\bm{\sigma}}

\newcommand{\Fermi}{{\textrm{F}}}
\newcommand{\Hydro}{{\textrm{H}}}
\newcommand{\sound}{{\textrm{s}}}
\newcommand{\throat}{{\textrm{t}}}
\newcommand{\sw}{{\textrm{sf}}}
\newcommand{\Ham}{{H}}
\newcommand{\Hawking}{{\textrm{H}}}

\newcommand{\marked}[1]{#1}

\begin{document}

\title{Voltage characteristics of hydrodynamic Dirac electron nozzles with supersonic flow}

\author{Kristof Moors}
\email{kristof.moors@imec.be}
\affiliation{Imec, Kapeldreef 75, 3001 Heverlee, Belgium}
\affiliation{Peter Grünberg Institute (PGI-9), Forschungszentrum Jülich, 52425 Jülich, Germany}
\affiliation{Jülich Aachen Research Alliance (JARA), Fundamentals of Future Information Technology, 52425 Jülich, Germany}
\affiliation{Instituut voor Theoretische Fysica, Department of Physics and Astronomy, KU Leuven, Celestijnenlaan 200D, 3001 Heverlee, Belgium}
\affiliation{Department of Physics and Materials Science, University of Luxembourg, 1511 Luxembourg, Luxembourg}
\author{Oleksiy Kashuba}
\affiliation{Institut f\"ur Theoretische Physik und Astrophysik, Universit\"at W\"urzburg, D-97074 W\"urzburg, Germany}
\author{Thomas L. Schmidt}
\affiliation{Department of Physics and Materials Science, University of Luxembourg, 1511 Luxembourg, Luxembourg}

\hyphenation{}

\date{\today}

\begin{abstract}
	In clean Dirac electron systems such as graphene, electron-electron interactions can dominate over other relaxation mechanisms such as phonon or impurity scattering. In this limit, collective electron dynamics can be described by hydrodynamic equations. The prerequisites for electron hydrodynamics have already been fulfilled in experiments, and signatures of hydrodynamic flow have been identified in transport measurements.
	Here, we derive the pressure-driven hydrodynamic flow profile across a de Laval nozzle profile for Dirac electrons in the subsonic and supersonic regimes. Based on this, we resolve the local voltage characteristics, which provide clear signatures of supersonic hydrodynamic flow.
	In particular, we identify two distinct features in the experimentally measurable potential profile: a pronounced asymmetry of the local voltage profile on opposite sides of the nozzle, and a sharp differential resistance signature induced by an electron shock wave on the exit side of the nozzle.
\end{abstract}

\maketitle

\section{Introduction}
\label{sec:introduction}
Various electronic transport phenomena can be traced back to the propagation of individual charge carriers, in a ballistic or diffusive regime, for example. The description as individual carriers provides an extremely versatile framework, as the electrons in many condensed matter systems are well described as almost free quasiparticles.
A very different transport regime, namely \emph{hydrodynamic} electron flow, takes over in the opposite limit of very strong interparticle interactions~\cite{Gurzhi1968, Narozhny2022}. Rather than relying on individual quasiparticles, modeling such transport is based on notions from the classical theory of hydrodynamics, such as the continuity equation and the Navier-Stokes equation. Hydrodynamic flow is possible irrespective of whether the underlying particles are fermionic~\cite{Gurzhi1963, Gurzhi1968} or bosonic~\cite{Gurzhi1964, Guyer1966, Shklovskii1967, Shklovskii1969a, Shklosvkii1969b, Eguiluz1976, Schwartz1982}.

However, reaching the regime of hydrodynamic electron flow in experiments has proved difficult: in most materials deviations from purely ballistic transport are either caused by disorder-induced scattering (for instance, due to impurities) at low temperatures, or by electron-phonon scattering at higher temperatures. Both of these scattering mechanisms drive the system to a diffusive transport regime and thus inhibit hydrodynamic electron flow. In recent years, ultra-pure two-dimensional materials have emerged as an ideal platform for reaching the hydrodynamic regime~\cite{Narozhny2022}, with graphene being one of the notable examples~\cite{Levitov2016, Lucas2018}. In sufficiently clean graphene samples, a large temperature window appears where electron-electron interactions dominate over disorder-induced scattering and electron-phonon interactions~\cite{Bandurin2018, Ho2018}. In this temperature range, hydrodynamic flow can be realized with distinct transport signatures.

Several effects have already been proposed and investigated as signatures of hydrodynamic behavior, including vortex formation with an associated negative nonlocal resistance signature~\cite{Torre2015, Levitov2016, Bandurin2016, Pellegrino2016, Shytov2018, GuerreroBecerra2019, AharonSteinberg2022, Palm2024}, viscous flow in the presence of boundaries (e.g., Poiseuille flow profile), barriers or constrictions~\cite{deJong1995, Kiselev2019, Sulpizio2019, Gusev2020, Ku2020, Vool2021, Jenkins2022, Krebs2023, Vijayakrishnan2025}, hydrodynamic thermoelectric behavior~\cite{Crossno2016, Ghahari2016, Gooth2018, Jaoui2018, Tikhonov2019}, the Gurzhi effect (superballistic transport)~\cite{Gurzhi1963, Guo2017, KrishnaKumar2017, Kashuba2018, Gusev2018a, Gusev2021, Ginzburg2021}, and viscous hydrodynamic magnetotransport (Hall viscosity)~\cite{Alekseev2016, Scaffidi2017, Pellegrino2017, Delacretaz2017, Gusev2018, Gusev2018, Berdyugin2019, Narozhny2019a}.

All the effects listed above appear in the regime of \textit{subsonic} incompressible hydrodynamic flow. However, when the flow speed becomes comparable to the speed of sound, an even richer phenomenology due to compressible hydrodynamic flow can be expected.
In classical systems, a de Laval nozzle is widely used for steam turbines and rocket or jet engines, and the underlying physics has numerous applications in other areas of physics. In particular, a relativistic de Laval nozzle provides a simple description of jets near black holes or neutron stars~\cite{Blandford1974, Rezzolla2013}. In a condensed-matter context, such nozzle geometries have been considered for the realization of sonic black holes~\cite{Novello2002}, e.g., in trapped Bose-Einstein condensates~\cite{Barcelo2003, Leonhardt2003, Jain2007}, with the analog of an event horizon appearing where the flow enters the supersonic regime.

Here, we propose graphene shaped into a nozzle geometry as a feasible experimental setup for the investigation of the compressible hydrodynamic regime with flow speeds approaching and even exceeding the speed of sound, i.e., realizing \emph{supersonic} flow.
For a hydrodynamic Dirac electron system such as graphene, we find that a de Laval nozzle displays a number of electronic transport features that can be taken as strong indicators of supersonic hydrodynamic transport. The main feature is an abrupt change in flow properties with the appearance of a shock front~\cite{Gardner1991}, as the flow across the nozzle transitions to supersonic speeds (the speed of sound of a two-dimensional hydrodynamic Dirac electron system is $v_\sound = v/\sqrt{2}$, where $v$ is the Dirac velocity~\cite{Landau1987, Rezzolla2013, Phan2013}).

The Article is structured as follows. We will present the hydrodynamic equations and the resulting equations that govern the nozzle flow in Sec.~\ref{sec:model}. In Sec.~\ref{sec:pressure-driven-flow}, we work out the pressure-driven flow profiles across the nozzle, and we derive the corresponding voltage characteristics in Sec.~\ref{sec:voltage-characteristics}. We discuss the underlying assumptions of our modeling approach and the resulting properties of hydrodynamic Dirac electrons in de Laval nozzles in Sec.~\ref{sec:discussion}, before concluding in Sec.~\ref{sec:conclusion}. Technical details, supporting results, and a list of symbols are provided in Appendixes~\ref{sec:Dirac_macro}-\ref{sec:notation}.

\begin{figure}
	\centering
	\includegraphics[width=\linewidth]{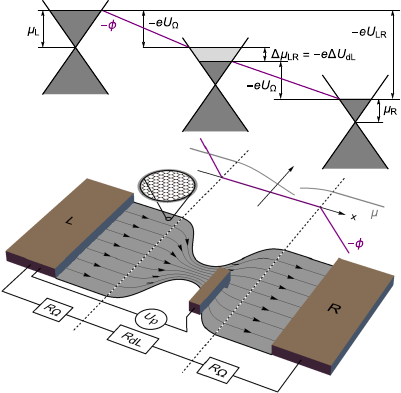}
	\caption{
		A graphene-based de Laval nozzle that is connected to two leads, over which a bias voltage $U_{\Le\Ri}$ is applied, and a noninvasive probe that locally measures the voltage difference with respect to the left lead $U_\mathrm{p}$. The lines with arrows indicate the laminar hydrodynamic flow of charge carriers through the nozzle.
        The lumped-element model for resolving the voltage characteristics, with Ohmic \marked{sections and a hydrodynamic constriction region} between two leads\marked{, and the corresponding chemical potential and electric potential profiles are shown below and above the device schematic, respectively}.
	}
	\label{fig:setup}
\end{figure}

\section{Model}
\label{sec:model}
\subsection{Hydrodynamic equations in Dirac systems}
\label{subsec:hydrodynamic-equations}
We consider massless Dirac fermions with the kinetic Hamiltonian $H(\vecp) = v \, \vecsigma \cdot \vecp$ in two ($D=2$) or three ($D=3$) spatial dimensions (in units with $\hbar = 1$), where $v$ is the Dirac velocity. For $D=2$, $\vecsigma = (\sigma_x, \sigma_y)^\intercal$ is the vector of Pauli matrices, and $\vecp= (p_x, p_y)^\intercal$ is the momentum (defined analogously for $D=3$).

In the limit of strong interparticle interactions (with the particles being Dirac electrons or holes), other interactions can be neglected (e.g., with impurities or phonons) and this system can be described by the momentum-conserving hydrodynamic equations of a nonviscous fluid~\cite{Rezzolla2013, Lucas2018}.
While being a viscous fluid in general, a hydrodynamic Dirac electron system can be described as a nonviscous fluid when expanding the hydrodynamic equations as a function of $l_{e\textnormal{-}e}/L$ and only keeping the zeroth order terms, with $l_{e\textnormal{-}e}$ the typical interparticle scattering length and $L$ the typical length scale of the flow profile (i.e., the length of the nozzle in our case). This approximation offers a good starting point for resolving the flow profile and allows us to solve the hydrodynamic equations in the nozzle analytically. For more details on the impact of viscosity in our setup, see the Discussion section (Sec.~\ref{sec:discussion}) and Appendix~\ref{sec:viscous_flow}. Furthermore, momentum-conserving hydrodynamic flow implies that $L$ is smaller than the typical length scale for momentum relaxation $l_\textnormal{mom}$, which is induced, for example, by collisions with impurities or phonons. Therefore, it is essential that the design of the nozzle satisfies the constraints $l_\textnormal{mom} > L > l_{e\textnormal{-}e}$, provided that such a window with hydrodynamic transport exists (see Sec.~\ref{sec:discussion}). For a very clean graphene sample at around $100\,\text{K}$, $L$ should be of the order of $1\,\text{\textmu m}$~\protect\cite{Lucas2018}.

Under the assumptions mentioned above, it is possible to define macroscopic quantities such as the charge carrier density $N(\vecr)$, (effective fluid) mass density $M(\vecr)$, energy density $E(\vecr)$, hydrodynamic pressure $P(\vecr)$, and the flow velocity $\vecV(\vecr)$ that satisfy the hydrodynamic Euler equations,
\begin{align} \label{eq:hydro_final}
\begin{split}
	\nabla P + M (\vecV \cdot \nabla) \vecV &= 0, \\
	\nabla \cdot (M \vecV) = 0, \quad \nabla \cdot (N \vecV) &= 0.
\end{split}
\end{align}
Their precise definitions and derivation from the quantum kinetic equation can be found in Appendix~\ref{sec:Dirac_macro}.
These Euler equations are, respectively, manifestations of the momentum, energy, and particle number conservation laws respected by the electron-electron interactions in the stationary (zero-frequency) regime.
Note that we neglect electric and magnetic fields in the Euler equations above. In typical hydrodynamic transport equations for a two-dimensional electron gas, \marked{the electric field, commonly related to the charge density gradient via the gradual-channel approximation~\cite{Dyakonov1993}, governs the flow.}
However, \marked{the electric field can be suppressed with sufficient screening or} the local charge density and electric potential can be decoupled through a tailored sample and backgate design, as recently demonstrated for a bilayer graphene de Laval nozzle in Ref.~\cite{Geurs2025}.
Here, we consider hydrodynamic flow across a de Laval nozzle that is driven predominantly by a pressure gradient, which can be induced by a small chemical potential or temperature difference across the nozzle (see Sec.~\ref{sec:voltage-characteristics} for more details).

The relation between the mass density $M$ (or enthalpy density $M v^2$) and the pressure $P$ of our relativistic system is given by
\begin{equation} \label{eq:PMV}
	P = M (v^2 - V^2) / (D + 1) = M v^2 - E,
\end{equation}
where $V = |\vecV|$, and the last equality relates the pressure to the energy density. A well-known result from relativistic hydrodynamics states that the speed of sound $v_\sound$ of a $D$-dimensional Dirac system is equal to $\sqrt{\deriv P/ \deriv E} = v/\sqrt{D}$~\cite{Rezzolla2013}, so supersonic flow corresponds to $V > v_\sound$. Note that the speed of sound $v_\sound$ refers to the propagation of pressure waves of the Dirac electron fluid~\cite{Lucas2016B}, not to be confused with the speed of sound (phonon dispersion) related to the crystal lattice of the host material (e.g., graphene).

\subsection{Nozzle equations}
\label{subsec:nozzle-equations}
Now, we apply the hydrodynamic equations that describe a strongly interacting electronic Dirac system driven by pressure in two $(D = 2)$ or three $(D = 3)$ spatial dimensions without momentum relaxation to a system with a nozzle geometry, as shown in Fig.~\ref{fig:setup}.
The nozzle is characterized by a varying cross section $A(x)$ (which has the dimension of length for the two-dimensional case and of an area for the three-dimensional case) as a function of the nozzle coordinate $x$, along which the flow is directed. We consider a smooth change of the cross section of the nozzle, i.e., $\deriv(A^{1/(D-1)})/\deriv x \ll 1$, and assume that the macroscopic quantities are uniform in the transverse directions (i.e., perpendicular to the flow direction).
Note that the cross section, in general, refers to the \textit{effective} cross section for the interior of the nozzle, where the fluid flows freely without direct influence from the boundaries. Near the boundaries, the flow speed may be reduced due to friction~\protect\cite{Erdmenger2018}, which would violate the assumption of uniformity of the flow profile along the transverse directions.
Turbulent flow would also violate the uniformity assumption, but is not expected in realistic samples in the regime dominated by electron-electron interactions. The Reynolds number $\mathcal{R}$ of the nozzle can be estimated by $\mathcal{R}\sim V L / (v \, l_{e\textnormal{-}e})$, with $V$ the flow velocity, $L$ the typical length scale of the flow profile (i.e., the length of the nozzle in our case), $v$ the Dirac velocity, and $l_{e\textnormal{-}e}$ the interparticle scattering length~\cite{Lucas2018}.
Turbulent flow is only expected for a Reynolds number of the order of $10^3$ or higher, requiring hydrodynamic transport over very large distances compared to the interparticle scattering length, which is typically prevented by momentum relaxation.

Under these assumptions, the flow profile is effectively one-dimensional~\cite{Courant1976} and the hydrodynamic equations simplify to:
\begin{align}
\label{eq:nozzle}
	\partial P + M V \partial V &= 0,& \partial ( M V A ) &= 0, & \partial(N V A) &= 0,
\end{align}
where $\partial \equiv \partial/\partial x$, $P$ is the pressure, $V$ the flow velocity, $M$ the effective fluid mass density, and $N$ the particle density.
The first equation is the stationary one-dimensional Navier-Stokes momentum equation in the nonviscous limit ($l_{e\textnormal{-}e} \rightarrow 0$).
The last two equations are continuity equations that reflect the conservation of particle current $I$ and momentum $S$, given by
\begin{align} \label{eq:conservation_laws}
	I &= N V A, & S &= M V A,
\end{align}
with $N$, $V$, and $A$ being functions of the nozzle coordinate $x$. The electrical current is given by $e I$, and the energy flow by $v^2 S$.
Here, we focus on the velocity profile along the flow direction of the nozzle. This is notably different from previous works that focus mainly on the velocity profile perpendicular to the flow direction of highly viscous hydrodynamic Dirac systems with constrictions~\cite{Guo2017, KrishnaKumar2017, Kiselev2019}.

Using Eq.~\eqref{eq:nozzle}, we can express the change in flow speed $V$ and pressure $P$ with the nozzle cross section $A$ as
\begin{align} \label{eq:nozzle_sol}
\begin{split}
	\frac{\partial V}{V} &= - \frac{1 - (V/v)^2}{1 - (V/v_\sound)^2} \frac{\partial A}{A}, \\
	\frac{\partial P}{P} &= \frac{(V/v)^2+(V/v_\sound)^2}{1 - (V/v_\sound)^2} \frac{\partial A}{A}.
\end{split}
\end{align}
These relations essentially govern the flow through a nozzle and we therefore refer to them as the \textit{nozzle equations}. Their derivation is provided in Appendix~\ref{sec:nozzle}.

The nozzle equations tell us that, if the flow starts at subsonic speed $V < v_\sound$, $V$ increases as the cross section decreases. This is a well-known consequence of Bernoulli's law. However, as soon as $V$ exceeds $v_\sound$, the behavior reverses and $V$ increases further with increasing cross section. This is the basic working principle of a de Laval nozzle: a section with decreasing cross section first accelerates the flow to the speed of sound, which is then attained at the throat of the nozzle (i.e., at the narrowest point with cross section $A_\throat$). Beyond the throat, an increasing cross section further accelerates the flow.

Solving for the flow speed as a function of the cross section with Eq.~\eqref{eq:nozzle_sol}, we obtain
\begin{equation} \label{eq:CA}
	\frac{C_A^2}{A^2} = \frac{V^2 (v^2 - V^2)^{D - 1}}{ v^{2D}} \leq \kappa^2 \equiv \frac{(D - 1)^{D-1}}{D^{D}},
\end{equation}
with integration constant $C_A$ (assumed to be positive without loss of generality). This constant fixes the relation between flow speed and cross section and can, up to a prefactor [see Eqs.~\eqref{eq:sol_subsonic}-\eqref{eq:sol_supersonic} in  Appendix~\ref{subsec:mu_and_T}], be thought of as the total particle (or electrical) current $I$ that flows through the nozzle.
Note that there is an upper bound $\kappa$ for $C_A/A$ and, hence, also for the current, which can only be reached when $V = v_\sound$ at the throat.
A solution for the pressure can also be obtained from Eq.~\eqref{eq:nozzle_sol} and is given by
\begin{equation} \label{eq:CP}
	\frac{C_A}{A} = \lef\frac{P}{C_P}\rig^{-\frac{D - 1}{D + 1}}
		\sqrt{1 - \lef\frac{P}{C_P}\rig^{-\frac{2}{D+1}}},
\end{equation}
where we have introduced the integration constant $C_P$, which corresponds to the pressure for vanishing flow speed.

\begin{figure}
	\centering
	\includegraphics[width=\linewidth]{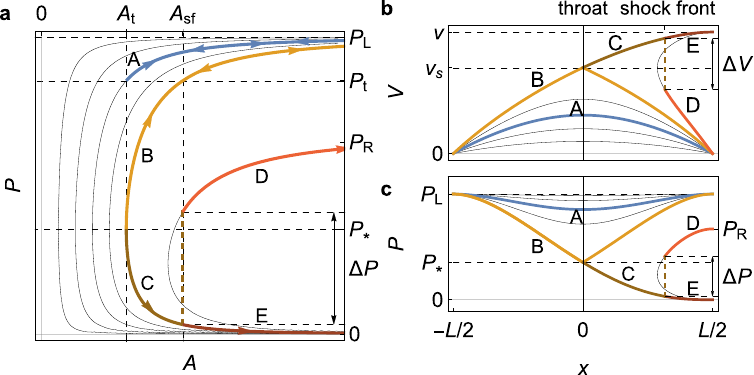}
	\caption{
		(\textbf{a}) The relation between pressure and cross section for the hydrodynamic flow through a two-dimensional nozzle. Three flow profiles are indicated: \textit{subsonic} flow with the pressure reaching pressure $P_\throat$ at the throat with cross section $A_\throat$ and returning to the initial pressure $P_\Le$ (blue line A, back and forth), \textit{critical} flow that reaches the critical pressure $P_*$ and the speed of sound at the throat before returning to the initial pressure (yellow line B, back and forth), and \textit{supersonic} flow with supersonic flow speeds between the throat and the shock front at cross section $A_\sw$ (line B, C \& D), where there is a pressure jump $\Delta P$ and a speed drop $\Delta V$ (brown dashed line). The \textit{ideal} supersonic flow profile is realized for $P_\Ri=0$ (line B, C \& E).
		(\textbf{b}),(\textbf{c}) The (\textbf{b}) flow speed and (\textbf{c}) pressure profiles as a function of the position along the nozzle are shown for the flow profiles indicated in (a) matching the corresponding labels and colors.
		We consider a two-dimensional nozzle with length $L$ and width profile given by $A(x) = A_\throat/[1-(2x/L)^2]$ here, such that the leads are infinitely wide: $A(\pm L/2) = +\infty$.
	}
	\label{fig:distribution}
\end{figure}

\section{Pressure-driven flow}
\label{sec:pressure-driven-flow}
To discuss the generic flow behavior of a Dirac electron fluid through a de Laval nozzle, it is convenient to consider a nozzle with length $L$ and nozzle coordinate $-L/2 \leq x \leq L/2$, attached to infinitely wide leads, i.e., $A(x = \pm L/2)^{1/(D - 1)} = +\infty$. Then, the possible boundary conditions for any flow profile are restricted to $V(x = \pm L/2) = 0,v$ [see Eq.~\eqref{eq:CA}], which is convenient to resolve the different flow profiles~\cite{Courant1976}.

Every solution of the nozzle equations [Eq.~\eqref{eq:nozzle_sol}] with a flow speed that remains subsonic along the length of the nozzle leads to equal pressure at the entrance and the exit (Fig.~\ref{fig:distribution}, line A), where the flow speed vanishes and the pressure is equal to $C_P$ [see Eq.~\eqref{eq:CP}].
The different \textit{subsonic} flow profiles correspond to different values of $C_A$ or, equivalently, the current, with $ 0 \leq C_A \leq \kappa A_\throat$. As $C_A$ increases, the maximal flow speed, which is realized at the throat and equal to zero when $C_A = 0$, increases until it reaches the speed of sound when $C_A = \kappa A_\throat$. This value of $C_A$ corresponds to the \emph{critical} flow profile shown in Fig.~\ref{fig:distribution} (line B), with the pressure dropping to the \textit{critical} pressure $P_* = C_P(1-1/D)^{(D+1)/2}$ at the throat and returning to the initial pressure at the nozzle entrance.
As we shall see in the following, a pressure gradient can be induced by a gradient of the chemical potential or temperature. Hence, in the subsonic regime, a finite current can flow with an infinitesimal bias voltage or heat gradient, up to the maximal current that is proportional to the cross section of the throat and $\kappa$.
When considering viscosity or momentum relaxation in realistic samples, there can be subsonic flow with finite pressure differences (bias voltages or heat gradients) between the leads and a smoother onset of supersonic flow is expected (see Appendix~\ref{sec:viscous_flow} and Ref.~\cite{Geurs2025}).

In addition to the \textit{critical} flow profile, there is an alternative solution of Eqs.~\eqref{eq:CA} and \eqref{eq:CP} with $C_A = \kappa A_\throat$, where the flow continues to accelerate, exceeding the speed of sound and reaching $V = v$ at the right lead, and the pressure continuing to decrease further beyond the throat (line C \& E). The solution is referred to as the \textit{ideal} flow profile and is realized when the pressure at the exit is equal to zero. We will see below that this requires the exit lead to be at zero temperature with chemical potential tuned precisely at the Dirac point, which is impossible to realize in practice.

Next, we consider two leads with different but finite pressures, denoted by $P_{\Le(\Ri)}$ for the left (right) lead, which necessarily induces a \textit{supersonic} flow profile. Without loss of generality, we assume $P_\Le > P_\Ri$, keeping $P_\Le$ fixed, such that the flow (of charge carriers) always goes from left to right.
Similarly to the solution for \textit{ideal} (supersonic) flow, the solution follows line B and C in Fig.~\ref{fig:distribution}. However, the flow must return to subsonic speeds to reach $V = 0$ at the right lead, and this implies that the nozzle equations become singular at a certain position past the throat [see Eq.~\eqref{eq:nozzle_sol} with $V = v_\sound$ and $\partial A/A \neq 0$], corresponding to a line (for $D=2$) across the nozzle. Therefore, the values of integration constants $C_A$ and $C_P$ need not be the same to the left and right of this position and we can obtain two different pressure profiles from Eq.~\eqref{eq:CP}, with $C_P^{\Le(\Ri)} = P_{\Le(\Ri)}$ for the solution that matches the pressure in the left (right) lead.
To the left of the position where the nozzle equations become singular, we have $C_A^\Le = \kappa A_\throat$, as for the \textit{critical} and \textit{ideal} flow profile.
To the right, the value of $C_A^\Ri$ follows from the conservation of momentum along the nozzle, yielding $C_P^\Le C_A^\Le = C_P^\Ri C_A^\Ri$ (see Appendix~\ref{subsec:pressure}), which, in turn, yields $C_A^\Ri = \kappa A_\throat P_\Ri / P_\Le$.
Having obtained $C_A$ and $C_P$ to the left and right of the singular point, one can see that there is a discontinuity in flow speed and pressure, as indicated by the dashed brown lines in Fig.~\ref{fig:distribution}. The latter reflects the appearance of a shock wave, which is a well-known feature of supersonic hydrodynamic flow profiles in de Laval nozzles~\cite{Courant1976}.

The shock front appears to the right of the throat and its position $x_\sw$ along the nozzle can be obtained by first integrating the hydrodynamic equations over the infinitesimal interval $\lim_{\delta \rightarrow 0}[x_\sw - \delta, x_\sw + \delta]$, then inserting the solutions for the pressure and velocity of the left and right limits, and finally solving for the cross section of the shock front $A_\sw$.
The second equality in Eq.~\eqref{eq:nozzle} yields $\Delta(MV) = 0$, with $\Delta X \equiv \lim_{\delta \rightarrow 0} \int_{x_\sw - \delta}^{x_\sw + \delta} \partial X \; \deriv x$ denoting a jump of the macroscopic quantity across the discontinuity, which can be used together with the first equality to obtain $\Delta P + M V \Delta V = 0$.
Inserting the expression for the pressure of Eq.~\eqref{eq:PMV}, we get the following condition for the discontinuity of the flow velocity:
\begin{equation}
	\Delta(v^2/V + DV)  = 0.
	\label{eq:sfpos}
\end{equation}
The numerical solution of this equation is given in Appendix~\ref{subsec:pressure}.
Starting from equal pressure and lowering the pressure in the right lead, a shock front appears near the throat and gradually shifts to the right lead, where it vanishes again. This is how the flow profile evolves from the \textit{critical} to the \textit{ideal} profile.

Note that the current flowing through the nozzle does not change for any supersonic profile between the \textit{critical} and \textit{ideal} profiles, as the current is determined by the pressure in the left lead, which is kept fixed. The current and flow speed saturate at their maximum value at the throat when reaching the sonic barrier and remain constant as the pressure in the right lead decreases.
Also note that the solution to the left of the shock front does not depend on the value of the pressure in the right lead. This is expected because the flow of information is bounded by the speed of sound of the Dirac fluid, and hence the regions are causally disconnected. It is the position of the shock front itself that shifts when varying the pressure in the right lead, along with a change in the flow profile to its right.
At the shock front, there is a pressure jump $\Delta P$, which, in the case of $D = 2$, is maximal and equal to $\Delta P_{\max} \approx 0.41 \, P_\Le$ when $P_\Ri \approx 0.81 \, P_L$, occurring at the position in the nozzle to the right of the throat where the cross section equals $A_\sw \approx 1.34 \, A_\throat$ (see Appendix~\ref{subsec:pressure} for details).

\section{Voltage characteristics}
\label{sec:voltage-characteristics}
So far, we have considered the flow through a nozzle in terms of the pressure, as in a conventional de Laval nozzle. However, since the temperature and chemical potential in the leads are the more accessible control parameters in electronic Dirac systems, we will study their effect on the flow profile in the following.

Based on explicit expressions for the particle number, mass density, and pressure in terms of the chemical potential, temperature, and flow speed of a hydrodynamic Dirac system (see Appendix~\ref{subsec:mu_and_T}), we obtain
\begin{align} \label{eq:N}
	N &\propto \frac{T^D v}{(v^2 - V^2)^{(D + 1)/2}} F_D^-(\mu/T), \\ \label{eq:M}
	M &\propto \frac{T^{D+1} v}{(v^2 - V^2)^{(D + 3)/2}} F_{D+1}^+(\mu/T) \propto \frac{P}{v^2 - V^2},
\end{align}
where $F_n^\pm(x) \equiv -[\Li_n(-\e^x) \pm \Li_n(-\e^{-x})]$ and $\Li_n(x)$ are polylogarithm functions, $T$ is the temperature, and $\mu$ is the chemical potential.
Rewriting Eq.~\eqref{eq:nozzle} in terms of temperature and chemical potential, we obtain
\begin{equation}
\label{eq:nozzle_3}
	\frac{\partial T}{T} = \frac{2 (V/v)^2}{1 - (V/v_\sound)^2} \frac{\partial A}{A} = \frac{\partial \mu}{\mu},
\end{equation}
with solution given by
\begin{equation}
\label{eq:CT}
	C_T / T = v^2 / (v^2 - V^2) = C_\mu / \mu,
\end{equation}
where, similar to $C_P$, the integration constants $C_T$ and $C_\mu$ represent the temperature and chemical potential, respectively, at vanishing flow speed.
We denote the chemical potential and temperature in the left (right) lead by $\mu_{\Le}$ $(\mu_\Ri)$ and $T_{\Le}$ ($T_\Ri$), respectively.

\begin{figure}[tb]
	\centering
	\includegraphics[width=\linewidth]{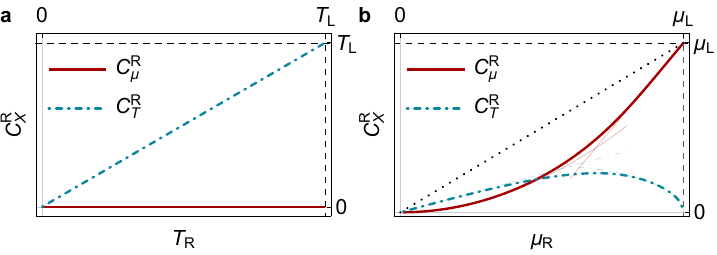}
	\caption{
		(\textbf{a}),(\textbf{b}) The chemical potential and temperature for vanishing flow speed at the nozzle exit to the right of the shock front in the supersonic regime, in the limit regimes with (\textbf{a}) $T_{\Le,\Ri} \gg \mu_{\Le,\Ri}$ and (\textbf{b}) $T_{\Le,\Ri} \ll \mu_{\Le,\Ri}$.
	}
	\label{fig:CX_TR_muR}
\end{figure}

We assume $\mu_{\Le,\Ri} > 0$ and low temperatures in comparison ($T_{\Le, \Ri} \ll \mu_{\Le, \Ri}$), such that the flow is induced by a chemical potential difference $\Delta \mu_\mathrm{LR} \equiv \mu_\Le - \mu_\Ri$, corresponding to a bias voltage $U_{\Le \Ri} = \Delta\mu_\mathrm{LR}/(-e)$ (see Appendix~\ref{subsec:mu_and_T} for details and for the opposite limit regime with $T_{\Le, \Ri} \gg \mu_{\Le, \Ri}$).
Note that experimental signatures of hydrodynamic flow have already been reported in the regime under consideration here~\cite{Lucas2018}.

From the explicit expression of the pressure in terms of the chemical potential in Eq.~\eqref{eq:M}, it follows that $P \propto \mu^{D + 1}$ in the low-temperature limit, such that a pressure gradient with supersonic flow from left to right is realized when $\mu_\Le > \mu_\Ri$.
In this case, the flow profile inherits the temperature and chemical potential of the left lead, i.e., $C_\mu^\Le = \mu_\Le$ and $C_T^\Le = T_\Le$.
Unlike for pressure, whose gradient directly drives the hydrodynamic flow, we cannot independently match the constants $C_{\mu,T}$ for temperature and chemical potential to the right of the shock front with their respective values in the right lead.
The values of temperature and chemical potential in the nozzle, downstream of the shock front, can be obtained by making use of current and momentum conservation, yielding
\begin{align} \label{eq:C_mu_R}
	C_\mu^\Ri &= \begin{cases}
		2 \mu_\Ri - \mu_\Le & (\mu_\Le - \mu_\Ri \ll \mu_\Le) \\
		(9 \zeta_3)^{1/3} \mu_\Ri^2 / (4 \ln 2 \, \mu_\Le) & (\mu_\Le \gg \mu_\Ri)
	\end{cases}, \\
	C_T^\Ri &= \begin{cases}
		\sqrt{3 (\mu_\Le - \mu_\Ri) \mu_\Ri}/\pi & (\mu_\Le - \mu_\Ri \ll \mu_\Le) \\
		\mu_\Ri / (9 \zeta_3)^{1/3} & (\mu_\Le \gg \mu_\Ri)
	\end{cases},
\end{align}
where $\zeta_3$ is the Ap\'ery constant. The results are visualized in Fig.~\ref{fig:CX_TR_muR}.
Note that, indeed, the chemical potential at the nozzle exit does not match with the right lead ($C_\mu^\Ri \neq \mu_\Ri$), unlike for the pressure. Moreover, despite a low temperature in the leads, the temperature of the fluid at the nozzle exit, $C_T^\Ri$, is not necessarily small compared to $C_\mu^\Ri$. The Dirac fluid heats up significantly by passing through the shock front, while no dissipation term is explicitly included in our nozzle equations.

Having worked out the integration constant for the chemical potential to the right of the shock front, as given in Eq.~\eqref{eq:C_mu_R}, we obtain a fully analytical description of the local chemical potential profile throughout a de Laval nozzle in both subsonic and supersonic regimes. We can thus resolve the corresponding voltage characteristics (see Appendix~\ref{subsec:voltage}).
For this, we consider a 2D (graphene-based) de Laval nozzle that is connected to source and drain leads. While momentum relaxation is not included in our analytical solutions of the de Laval nozzle, Ohmic dissipation is hard to avoid in realistic experimental transport setups. To account for Ohmic dissipation, we consider a setup in which \marked{the constriction geometry of the de Laval nozzle, for which we have obtained an analytical description, is placed between two sections with constant width (see Fig.~\ref{fig:setup}), resembling the device geometry of Ref.~\cite{Geurs2025}}. Essentially, we are treating the \marked{drop in electric potential across the contacted device} through lumped elements (two Ohmic resistors with resistances $R_\Omega^\Le$ and $R_\Omega^\Ri$ to the left and right of the nozzle \marked{constriction}, respectively, determined by their dimensions as well as the mobility and carrier density of the sample) while keeping it separate from the \marked{(pressure-driven)} hydrodynamic flow across the constriction geometry itself. To obtain consistent solutions, we match the chemical potentials of the Ohmic sections with those at the ends of the nozzle, and match the current flowing through each element of the circuit for different bias voltages, while keeping the average chemical potential $\langle \mu \rangle = (\mu_\Le + \mu_\Ri)/2$ fixed (determined by the charge carrier density). This allows us to apply our analytical solutions and qualitatively resolve the impact of subsonic and supersonic flow profiles in the nozzle on the (local) voltage characteristics. The details of this approach are provided in Appendix~\ref{subsec:nozzle-Ohmic}.

In the subsonic regime with flow speed at the nozzle throat below the speed of sound, there is no chemical potential difference (so no finite voltage difference) over the \marked{constriction region} and the current-voltage relation is purely Ohmic: $U_\mathrm{LR} = (R_\Omega^\Le + R_\Omega^\Ri) I$. This subsonic regime is maintained up to a critical bias voltage for which the flow speed reaches the speed of sound at the throat of the nozzle. At higher bias voltages, the nozzle enters the supersonic regime and the bias voltage $U_\mathrm{LR}$ is split over the Ohmic sections and a voltage $U_\mathrm{dL}$ across the de Laval nozzle (see schematic in Fig.~\ref{fig:setup}): $U_\mathrm{LR} = (R_\Omega^\Le + R_\Omega^\Ri) I + U_\mathrm{dL}$, with $U_\mathrm{dL} = \Delta \mu_\mathrm{LR}/(-e)$.
Note that $\Delta \mu_\mathrm{LR} \ll \mu_\mathrm{L}, \mu_\mathrm{R} \ll |e U_\mathrm{LR}|$ for a realistic setup.

\begin{figure}[tb]
	\centering
	\includegraphics[width=\linewidth]{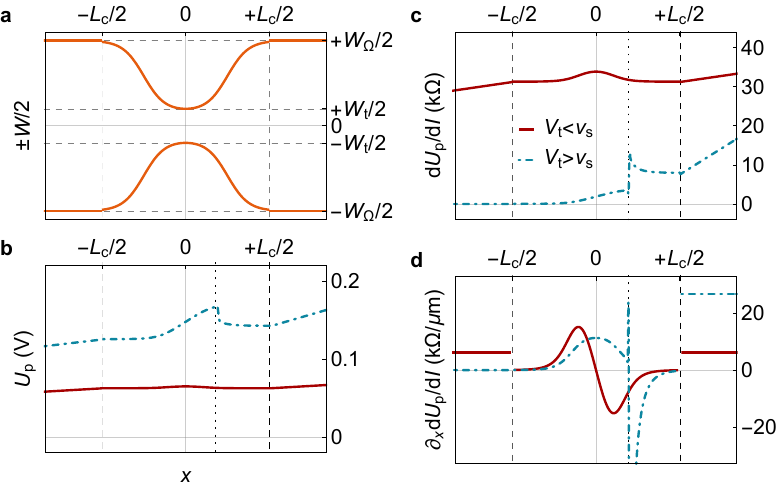}
	\caption{
        Voltage characteristics of a graphene-based de Laval nozzle with \marked{constriction geometry between two Ohmic sections with constant width}, as presented in Fig.~\ref{fig:setup}, in the subsonic ($V < v_\mathrm{s}$) and supersonic ($V > v_\mathrm{s}$) flow regimes.
		(\textbf{a}) Cross-sectional profile of the de Laval nozzle.
        (\textbf{b})-(\textbf{d}) The normalized (\textbf{a}) probe voltage, (\textbf{c}) differential resistance, and (\textbf{d}) its spatial derivative as a function of the position of the probe along the nozzle.
        The position of the shock front for the supersonic solution is indicated with a vertical dashed line.
        We consider graphene with $v =10^6\,\text{m/s}$, carrier density $n_\mathrm{2D} = 10^{11}\,\text{cm$^{-2}$}$, mobility $\mu_\mathrm{mob} = 10 000 \, \text{cm$^2$/(V$\cdot$s)}$, and a transport geometry with $5\,\text{\textmu m}$-long, $1\,\text{\textmu m}$-wide Ohmic sections and a $1\,\text{\textmu m}$-long \marked{constriction section} that is $0.2\,\text{\textmu m}$ wide at the throat (see Appendix~\ref{subsec:nozzle-Ohmic} for details).
	}
	\label{fig:voltage-characteristics}
\end{figure}

In Fig.~\ref{fig:voltage-characteristics}, we present the voltage characteristics of the graphene-based de Laval nozzle transport geometry shown in Fig.~\ref{fig:setup}. We consider the probe voltage $U_\mathrm{p}(x) \equiv R_\Omega^\Le I + [\mu_\mathrm{L} - \mu(x)]/(-e)$, which evaluates the voltage difference with respect to the left contact as a function of the position along the transport geometry. Note that, in the Ohmic sections, the potential drops linearly, as indicated in purple in Fig.~\ref{fig:setup}. When the subsonic and supersonic probe voltage profiles are compared, a clear difference in symmetry with respect to the throat of the nozzle can be observed. In the supersonic regime, the probe voltage increases further beyond the throat and drops sharply at the shock front position. This qualitative difference is even more pronounced for the local differential resistance $\mathrm{d}U_\mathrm{p}(x)/\mathrm{d}I$ and its spatial derivative, with the latter showing a pronounced peak where the shock front is positioned in the case of supersonic flow.

\section{Discussion}
\label{sec:discussion}
There are certain assumptions underlying our analytical modeling approach and we discuss them below.
First, we have assumed throughout this text that the Dirac electron fluid is nonviscous while, in real electron hydrodynamic systems, the interparticle scattering length is finite and the fluid therefore viscous~\cite{Sukhachov2021}.
A finite viscosity corresponds to the consideration of a finite interparticle scattering length when deriving the hydrodynamic equations from the quantum kinetic equation (see Appendix~\ref{sec:Dirac_macro}), giving rise to a viscosity term in the Navier-Stokes equation~\cite{Lucas2018}.
In Appendix~\ref{sec:viscous_flow}, we discuss in detail the impact of viscosity on the nozzle equations and the resulting flow profiles. In general, we find that an effective viscosity parameter governs the corrections to the 1D flow profiles, and these corrections become very small when the interparticle scattering length of the Dirac electrons is small compared to the dimensions of the nozzle.
For graphene, this scattering length can be on the order of $\sim 100\,\text{nm}$.
For sufficiently large nozzle dimensions, excellent quantitative agreement can be obtained between the viscous flow profiles (with low effective viscosity) and the nonviscous flow profiles (obtained in the perfect-fluid regime). The discontinuity in the flow profile turns into a continuous shock front remnant with a steep drop in flow speed, and, correspondingly, a steep upturn in pressure.
Hence, we expect the local voltage characteristics, as shown in Fig.~\ref{fig:voltage-characteristics}, to remain valid when the effective viscosity is sufficiently low, although the sharp (discontinuous) features related to the shock front would naturally get broadened by the viscosity.

Second, we consider hydrodynamic flow throughout \marked{the constriction region of} the nozzle without momentum relaxation, with Ohmic dissipation applied only in \marked{spatially separated} sections. As a consequence, any finite bias voltage or pressure difference over the \marked{constriction section} itself corresponds to a supersonic flow profile.
In Ref.~\cite{Geurs2025}, the flow profile of a bilayer graphene-based nozzle was resolved numerically, considering a linear potential profile throughout the \marked{constriction region}, while also considering viscous flow.
\marked{It is reassuring to see that, with} this approach, the resulting voltage characteristics in the supersonic regime are qualitatively similar to those obtained with our analytical model \marked{that fully neglects the Ohmics in the constriction region and considers pressure-driven hydrodynamic flow of a perfect fluid}. In particular, they also recover a clear peak in the profile of $\partial_x \mathrm{d}U_\mathrm{p}(x)/\mathrm{d} I$ downstream with respect to the nozzle throat, which is related to the presence of a shock front.
Note that in Ref.~\cite{Geurs2025} bilayer graphene is considered, which has a quadratic dispersion relation. Hence, the nozzle equations are only equal up to a relativistic factor. The speed of sound in bilayer graphene depends on the position of the Fermi level and is therefore gate-tunable. Nevertheless, similar voltage characteristics are obtained with a speed of sound that is of the same order of magnitude as that of single-layer graphene under consideration here.

Graphene seems to be a very promising candidate for the realization of supersonic hydrodynamic Dirac electron flow across a de Laval nozzle geometry. Large flow speeds (\mbox{$V > 0.1 \, v$}) and low electron densities, \mbox{$\sim  10^{11}\,\text{cm}^{-2}$}, have already been obtained with existing fabrication techniques and sample qualities~\cite{Meric2008, Dorgan2010, Yamoah2017}. The nozzle geometry itself should induce a further increase in speed so $V \approx v_\sound \approx 0.7 \, v$ seems to be within reach. Instead of using voltage probes, one could also verify supersonic flow directly by resolving the flow velocity profile past the throat of a graphene nozzle~\cite{Ella2019,Sulpizio2019}.
In addition to graphene, a Dirac de Laval-nozzle and its phenomenology can also be considered for other condensed matter systems with ($D = 2$ or $D = 3$) Dirac fermions, with the surface states of a 3D topological insulator and Dirac or Weyl semimetals as notable examples~\cite{Lucas2016, Galitski2018}.

Finally, we comment on how the sonic analog of a black hole can be realized with such a supersonic Dirac fluid, the region where the fluid turns supersonic representing the event horizon~\cite{Unruh1981}. The spread of information is bounded by the speed of sound in place of the speed of light in this hydrodynamic system. For a supersonic de Laval nozzle as considered here, quantized density waves or phonons of the hydrodynamic Dirac system are expected to be emitted from the throat of a supersonic nozzle toward the entrance with a black body spectrum, analogous to Hawking radiation forming near the event horizon of a black hole~\cite{Novello2002}. The Hawking temperature of this spectrum can be obtained from the flow speed through $T_\Hawking = \bigl. \partial(|V|-v_\sound) / (2\pi\kB)\bigr|_{|V|=v_\sound} = v / (2 \pi \kB L)$, with the last step obtained for the width profile considered in Fig.~\ref{fig:distribution}. The expression yields a temperature of the order of $1\,\text{K}$ for a graphene nozzle with length in the \textmu m range, comparable to the temperature of black hole analogs based on the hydrodynamic flow of microcavity polaritons~\cite{Nguyen2015}. It is the equivalent of a black hole with a mass one thousand times smaller than the mass of the earth. Although being two orders of magnitude lower than the typical temperature that is required in graphene to realize hydrodynamic transport, this Hawking temperature is rather high compared to other condensed-matter systems that have been proposed, such as superfluid helium or Bose-Einstein condensates, only yielding temperatures in the \textmu K~\cite{Jacobson1998} or nK~\cite{Barcelo2003} range.
Detection of this Hawking radiation can be envisioned with a very sensitive voltage probe that identifies the voltage fluctuations due to fluctuations in the fluid of Dirac electrons, and cross-correlating the fluctuations on opposite sides of the shock front would be able to disentangle the Hawking radiation from intrinsic temperature-induced fluctuations.
Optimizing the ratio of Hawking temperature versus the temperature of the Dirac electrons is crucial for the detectability of Hawking radiation. This is challenging, as lowering the temperature of the Dirac electrons will increase the interparticle collision length (typically, $l_{e\textnormal{-}e} \propto T^{-2}$), in turn limiting the minimal size of the nozzle and the maximal Hawking temperature that can be achieved.
Whether it could be observed in a given hydrodynamic Dirac system will ultimately depend on the details of the Dirac spectrum and the different scattering processes in that system (interparticle and momentum relaxing).

\section{Conclusion}
\label{sec:conclusion}
We have considered a de Laval nozzle to study the hydrodynamic behavior of strongly interacting Dirac electrons in condensed-matter systems such as graphene.
We consider \marked{pressure-driven hydrodynamic flow across a constriction geometry}, which can be realized with temperature or chemical potential gradients.
From the Euler equations for the hydrodynamic Dirac system, we derive the hydrodynamic flow profile across a nozzle in subsonic and supersonic regimes, with a shock wave being induced in the case of the latter.
This results in distinct voltage characteristics when applying a bias voltage between two leads at opposite ends of the nozzle, which can be resolved with a noninvasive local voltage probe.
Our findings suggest two distinctive voltage signatures related to hydrodynamic flow of Dirac electrons through a de Laval nozzle: a pronounced asymmetry of the local voltage profile on opposite sides of the nozzle when entering the supersonic regime and a sharp differential resistance signature related to an electron shock wave.

\section*{Acknowledgments}
The authors would like to thank Christophe De Beule, Patrik Recher and Johannes Geurs for fruitful discussions. K.M.\ and T.L.S.\ acknowledge the support by the National Research Fund Luxembourg with ATTRACT Grant No.~7556175 and O.K.\ acknowledges the support from SFB1170, ``ToCoTronics''.

\appendix
\renewcommand{\thefigure}{A\arabic{figure}}
\setcounter{figure}{0}

\section{Macroscopic quantities \& Hydrodynamic equations of Dirac systems}
\label{sec:Dirac_macro}
The hydrodynamic description of a $D$-dimensional ($D = 2$ or $D = 3$) Dirac system is based on the following macroscopic quantities: the particle number $N$, the current $\vecj$, the macroscopic momentum $\vecS$, macroscopic energy $E$, and the stress tensor $\rttensor{\Pi}$. They are defined as a function of the (semiclassical) electron distribution function $f_\lambda(\vecr, \vecp, t)$ as follows (with $\hbar = 1$):
\begin{align} \label{eq:N_apx}
\begin{split}
	N &= \sum_{\lambda, \vecp} f_\lambda (\vecp), \quad \vecj = \sum_{\lambda, \vecp} v \vecn f_\lambda (\vecp), \\
    \vecS &= \sum_{\lambda, \vecp} \vecp f_\lambda (\vecp), \quad E = \sum_{\lambda, \vecp} \lambda v p f_\lambda (\vecp), \\
	\rttensor{\Pi}_{ij} &= \sum_{\lambda, \vecp} v n_i p_j f_\lambda (\vecp),
\end{split}
\end{align}
where $v$ is the Dirac velocity, $\vecp$ is the momentum ($p \equiv |\vecp|$) and $\lambda = \pm 1$ the electron or hole nature of the state or, equivalently, its chirality, such that a state with momentum $\vecp$ and chirality $\lambda$ has an energy $E_\lambda(\vecp) = \lambda v p$. Moreover, $\vecn \equiv \lambda \vecp / p$ is a unit vector in the direction of the momentum. Here, we consider a stationary flow, in which case all macroscopic quantities will depend only on position $\vecr$ and not on time.
In addition, we will consider the macroscopic chirality $\Lambda$ and the chiral current $\vecj_\Lambda$, given by
\begin{align}
	\Lambda &= \sum_{\lambda, \vecp} \lambda f_\lambda(\vecp), &
	\vecj_\Lambda &= \sum_{\lambda, \vecp} \lambda v \vecn f_\lambda(\vecp).
\end{align}
We consider a Dirac system subject to interparticle collisions that conserve the total particle number, chirality, momentum, and energy, which can be represented by their intensive thermodynamic conjugate variables $\phi$, $\chi$, $\vecalpha$, and $\beta$, respectively.
The system can then be represented by a distribution function $f_\Hydro(\vecalpha,\beta,\chi,\phi) = f_\Fermi(\vecalpha \cdot \vecp + \beta \lambda v p + \chi \lambda + \phi)$, which cannot be affected by the interparticle collisions and can be expressed in terms of the Fermi-Dirac distribution $f_\Fermi(z) = 1/(\e^z + 1)$. We refer to $f_\Hydro$ as the \textit{hydrodynamic flow distribution function}, and proceed with the natural redefinition of $\vecalpha$ in terms of the flow velocity $\vecV = - \vecalpha / \beta$, of $\beta$ in terms of temperature $T = 1/\beta$, of $\phi$ in terms of the chemical potential $\mu = - \phi / \beta$, and of $\chi$ in terms of a chirality-dependent shift of the chemical potential $X = - \chi / \beta$ and  (with $\kB = 1$).
The particle number $N$, for example, can be obtained from the straightforward integration of the Fermi-Dirac distribution function as follows:
\begin{widetext}
\begin{equation}
	\begin{split}
		N &= \frac{S^{D-1}}{(2 \pi)^D} \int\limits_0^\pi \deriv\theta \; (\sin \theta)^{D-2} \! \int\limits_0^{+\infty} \! \deriv p
		\lef \frac{p^{D-1}}{\e^{[(v - V \cos\theta) p - (\mu + X)]/T} + 1} - \frac{p^{D-1}}{\e^{[(v + V \cos\theta) p + (\mu - X)]/T} + 1} \rig \\
		&= \frac{S^{D-1}}{(2 \pi)^D} \int\limits_0^\pi \deriv\theta \; (\sin \theta)^{D-2}
		\lef - \frac{\Gamma(D) T^D \Li_D(- \e^{(\mu + X)/T})}{(v - V \cos\theta)^D} + \frac{\Gamma(D) T^D \Li_D(- \e^{- (\mu - X)/T})}{(v + V \cos\theta)^D} \rig \\
		&= \frac{S^D}{(2 \pi)^D} \Gamma(D) \frac{T^D v}{(v^2 - V^2)^{(D + 1)/2}} F_D^-(\mu/T, X/T),
	\end{split}
\end{equation}
with gamma function $\Gamma(x)$, polylogarithm functions $\Li_n(x)$, and where we have made use of the relation $f_\Fermi(z) - 1 = - f_\Fermi(-z)$ and redefined $f_-(\vecp) \rightarrow f_-(\vecp) - 1$ to make the integral over momenta finite. The surface of a $D$-dimensional sphere $S^D$ and the function $F_D^-(\mu/T, X/T)$ were also introduced, defined as:
\begin{align}
	S^D &= \frac{2 \pi^{D/2}}{\Gamma(D/2)} = \begin{cases} 2 \pi \quad (D = 2) \\ 4 \pi \quad (D = 3) \end{cases} , \\ \label{eq:poly_apx}
	F_D^\pm(\mu/T, X/T) &\equiv - [\Li_D(-\e^{(\mu + X)/T}) \pm \Li_D(-\e^{-(\mu - X)/T})].
\end{align}
We can confirm this result by exploiting Lorentz invariance. We consider a Lorentz boosted reference frame with boost speed $V$ and momentum $\vecp'$, related to $\vecp$ as follows:
\begin{align}
	\begin{aligned}
    p'_\parallel &= \gamma (p_\parallel - \Upsilon |\vecp|), \\
    p_\parallel &= \gamma (p_\parallel' + \Upsilon |\vecp'|),
    \end{aligned} &&
	\begin{aligned}
    |\vecp'| &= \gamma (|\vecp| - \Upsilon p_\parallel), \\
    |\vecp| &= \gamma(|\vecp'| + \Upsilon p_\parallel'),
    \end{aligned} &&
    \begin{aligned}
	\vecp_\perp' &= \vecp_\perp, \\
    \int\frac{\deriv^D \vecp}{|\vecp|} &= \int\frac{\deriv^D \vecp'}{|\vecp'|},
    \end{aligned}
\end{align}
with $\Upsilon = V/v$, $\gamma = 1/\sqrt{1 - \Upsilon^2}$ and $p_\parallel$ ($\vecp_\perp$) the component(s) of the momentum parallel (perpendicular) to the boost direction. The last equation presents the Lorentz invariant integration measure over all momenta.
This can be used to obtain
\begin{equation}
	\begin{split}
		N &= \frac{1}{(2 \pi)^D} \int\deriv^D \vecp \; \left\{ f_\Fermi[(v |\vecp| - \vecV \cdot \vecp - \mu - X)/T] - f_\Fermi[(v |\vecp| + \vecV \cdot \vecp + \mu - X)/T] \right\} \\
		&= \frac{1}{(2 \pi)^D} \int\frac{\deriv^D \vecp'}{|\vecp'|}  \;
		\left\{ f_\Fermi[(v |\vecp'| / \gamma - \mu - X)/T] \gamma (|\vecp'| - \vecV \cdot \vecp') - (\mu \leftrightarrow - \mu, \vecV \leftrightarrow -\vecV) \right\} \\
		&= \frac{1}{(2 \pi)^D} \gamma^{1 + D} \int \deriv^D \vecp' \;
		\left\{ f_\Fermi[(v |\vecp'| - \mu - X)/T] - f_\Fermi[(v |\vecp'| + \mu - X)/T]\right\} \\
		&= \frac{S^D}{(2 \pi)^D} \gamma^{1 + D} \int\limits_0^{+\infty} \deriv p \;
		\lef \frac{p^{D-1}}{\e^{(v p - \mu - X)/T} + 1} - \frac{p^{D-1}}{\e^{(v p + \mu - X)/T} + 1} \rig \\
		&= \frac{S^D}{(2 \pi)^D} \Gamma(D) \frac{T^D \gamma^{1 + D}}{v^D} F_D^-(\mu/T, X/T),
	\end{split}
\end{equation}
where we have considered a boosted reference frame along the flow, in opposite directions for both terms.

Having obtained the other macroscopic quantities in a similar manner, one can verify that the following relations hold:
\begin{align} \label{eq:macrovalues_eq}
	\begin{aligned}
		\vecj_\Hydro &= N \vecV, \\
		E &= M v^2 - P = \Tr \, \rttensor{\Pi}_{(\Hydro)},
	\end{aligned} &&
	\begin{aligned}
		\vecj_{\Lambda \, \Hydro} &= \Lambda \vecV, \\
		P &= M (v^2 - V^2) / (D + 1).
	\end{aligned} &&
	\begin{aligned}
		\vecS &= M \vecV, \\
		&
	\end{aligned} &&
	\begin{aligned}
		\rttensor{\Pi}_\Hydro &= P + \vecS \otimes \vecV, \\
		&
	\end{aligned}
\end{align}
We have added a subscript '$\Hydro$' to the (chiral) current and the stress tensor, as these quantities are obtained from the hydrodynamic flow distribution function, but are not conserved by interparticle collisions. However, the relation between the energy and the trace of the stress tensor is valid in general.
Note that we have introduced the pressure $P$ as the component of the stress tensor for vanishing flow velocity, which can be shown to agree with the thermodynamic definition as the derivative of the energy with respect to the system volume for constant entropy and particle number \cite{Lucas2018}.
We have also introduced the effective fluid mass density $M$ that relates the flow velocity to the macroscopic momentum. It can be obtained in a similar manner as the particle number, yielding
\begin{equation} \label{eq:mass}
	M = \frac{S^D}{(2 \pi)^D} \frac{D + 1}{D} \Gamma(D + 1) \frac{T^{D+1} v}{(v^2 - V^2)^{(D + 3)/2}} F_{D + 1}^+(\mu/T, X/T).
\end{equation}
It is the analog of the mass density of a conventional fluid.

In this work, we do not consider chiral symmetry breaking, which would correspond to $X \neq 0$. This quantity only appears inside the functions $F_n^\pm(\mu/T,X/T)$, which can be expanded for small $X$ as:
\begin{equation}
	F_n^\pm(\mu/T,X/T) \approx F_n^\pm(\mu/T,0) + F_{n - 1}^\pm (\mu/T,0) X/T + \frac{1}{2} F_{n-2}^\pm(\mu/T,0) X^2/T^2.
\end{equation}
Hence, we have only considered the leading-order contribution.
This chiral symmetry is equivalent to considering an electron-hole-symmetric system, with the distribution for electrons and holes identical to each other upon changing the sign for energy and momentum.

Approximations for the macroscopic quantities can be obtained in the low- and high temperature regimes by making use of the following expansions:
\begin{align} \label{eq:F_exp_1}
	F_n^+(x) &= \begin{cases}
		2 (1-2^{1-n}) \zeta_n + (1 - 2^{3 - n})\zeta_{n - 2} x^2 & (x \ll 1, n \neq 3) \\
		3 \zeta_3/2 + \ln 2 \, x^2 & (x \ll 1, n = 3) \\
		\left[ |x|^n + \pi^2 n (n - 1) |x|^{n - 2} / 6 \right] / \Gamma(n + 1) & (x \gg 1)
	\end{cases} , \\ \label{eq:F_exp_2}
	F_n^-(x) &= \begin{cases}
		2(1-2^{2 - n})\zeta_{n-1} x & (x \ll 1, n \neq 2) \\
		2 \ln 2 \, x & (x \ll 1, n = 2) \\
		\left[ |x|^n + \pi^2 n (n - 1) |x|^{n - 2} / 6 \right] (\sign \, x) / \Gamma(n + 1) & (x \gg 1)
	\end{cases} ,
\end{align}
Here $F_n^\pm(x) \equiv F_n^\pm(x, 0)$, with the definition of Eq.~\eqref{eq:poly_apx}, and $\zeta_s = \sum_{k=1}^{\infty}k^{-s}$, with $\zeta_0 = -1/2$, $\zeta_2 = \pi^2/6$, $\zeta_3 \approx 1.20$, and $\zeta_4 = \pi^4/90$, for example. Note that a separate treatment is required for $F_3^+$ and $F_2^-$ as $\zeta_1 = +\infty$.
For the particle number and effective fluid mass density, for example, we obtain the following limits in two and three spatial dimensions:
\begin{align} \label{eq:D2N}
	D = 2: \quad N &\approx \frac{1}{2 \pi} \frac{v}{(v^2 - V^2)^{3/2}} \times
	\begin{cases} 2 \ln 2 \, T^2 (\mu/T) & (\mu \ll T) \\ (\sign \, \mu) \mu^2 / 2 & (\mu \gg T) \end{cases}, \\ \label{eq:D2M}
	M &\approx \frac{3}{2 \pi} \frac{v}{(v^2 - V^2)^{5/2}} \times
	\begin{cases} 3 \zeta_3 T^3 / 2 & (\mu \ll T) \\ |\mu|^3 / 6 & (\mu \gg T) \end{cases}, \\ \label{eq:D3N}
	D = 3: \quad N &\approx \frac{1}{\pi^2} \frac{v}{(v^2 - V^2)^2} \times
	\begin{cases} \pi^2 T^3 (\mu / T) / 6 (\mu \ll T) \\ \mu^3 / 6 (\mu \gg T) \end{cases}, \\ \label{eq:D3M}
	M &\approx \frac{4}{\pi^2} \frac{v}{(v^2 - V^2)^3} \times
	\begin{cases} 315 \pi^4 T^4 / 2 & (\mu \ll T) \\ |\mu|^4 / 24 & (\mu \gg T) \end{cases}.
\end{align}
\end{widetext}

The dynamics of the macroscopic quantities can be obtained from the semiclassical Boltzmann equation, which incorporates the scattering mechanisms through the collision integral \cite{Lifshitz1981}. We only consider the regime in which the interparticle collisions ($e$-$e$) are dominant, neglecting any other scattering mechanism:
\begin{equation}
	\partial_t f + v \vecn \cdot \nabla f -e \lef \vecE + \frac{v}{c} \vecn \times \vecB \rig \cdot \nabla_\vecp f = \mathcal{I}_{e\textnormal{-}e}[f],
\end{equation}
with drift term due to external electric and magnetic fields, $\vecE$ and $\vecB$, respectively, and collision integral $\mathcal{I}_{e\textnormal{-}e}[f]$.
From this equation, we obtain the following hydrodynamic equations for the particle number, chirality, momentum, and energy, noting that the right-hand side vanishes for these quantities:
\begin{align}
	\partial_t N + \nabla \cdot \vecj &= 0, \\
	\partial_t \Lambda + \nabla \cdot \vecj_\Lambda &= 0, \\ \label{eq:hydro_apx_3}
	\partial_t \vecS + \nabla \cdot \rttensor{\Pi} + e \vecE N + \frac{e}{c} \vecj \times \vecB &= 0, \\ \label{eq:hydro_apx_4}
	\partial_t E + v^2 \nabla \cdot \vecS + e \vecE \cdot \vecj &= 0,
\end{align}
where it is understood that the divergence on the third line acts on the first index of the stress tensor. Note that the flow of energy is proportional to the momentum in the absence of an electric field. They are related by a factor of $v^2$, as can be seen in Eq.~\eqref{eq:hydro_apx_4}.

Close to a hydrodynamic flow distribution, one can write $\vecj_{(\Lambda)} = \vecj_{(\Lambda) \, \Hydro} + \delta \vecj_{(\Lambda)}$ and $\rttensor{\Pi} = \rttensor{\Pi}_\Hydro + \delta \rttensor{\Pi}$, with small corrections $\delta \vecj_{(\Lambda)}$ and $\delta \rttensor{\Pi}$. The corrections can be obtained from the Boltzmann equation, linearized around the hydrodynamic flow distribution. We further assume the relaxation time approximation with collisions characterized by a single interparticle collision time $\tau_{e\textnormal{-}e}$ (the Callaway ansatz \cite{Callaway1959, DeGennaro1984, DeGennaro1985}), yielding:
\begin{equation} \label{eq:Boltzmann}
	\partial_t f + v \vecn \cdot \nabla f - e \lef \vecE + \frac{v}{c} \vecn \times \vecB \rig \cdot \nabla_\vecp f = - \frac{\delta f}{\tau_{e\textnormal{-}e}},
\end{equation}
with distribution function $f = f_\Hydro + \delta f$.
From this equation, it is clear that the corrections to the Fermi-Dirac values vanish in the nonviscous-fluid limit $\tau_{e\textnormal{-}e} \rightarrow 0$ for infinitely strong interparticle collisions.
It is important to note that the gradient terms vanish if $f_\Hydro$ is a position-independent function of the quantities. In this work, we mainly consider a space-dependent distribution function $f_\Fermi[v p/T(\vecr) - \vecV(\vecr)\cdot\vecp/T(\vecr) -\mu(\vecr)/T(\vecr)]$ whose distribution is captured by \textit{local} conjugate variables, according to the zeroth order approximation \cite{Rezzolla2013, Lucas2018}, being a suitable ansatz for resolving a flow profile that varies over length scales much larger than the interparticle scattering length.

In the stationary regime and in the absence of electric and magnetic fields, the hydrodynamic equations for particle number, momentum, and energy that follow from these considerations are given by:
\begin{align} \label{eq:hydro_final_apx}
\begin{split}
	\nabla \cdot \vecj_\Hydro &= 0 , \quad \nabla \cdot \vecj_{\Lambda \, \Hydro} = 0, \\
	\nabla \cdot \rttensor{\Pi}_\Hydro &= 0, \quad \nabla \cdot \vecS = 0.
\end{split}
\end{align}
Inserting the Fermi-Dirac relations of Eq.~\eqref{eq:macrovalues_eq}, we obtain precisely the hydrodynamic equations of Eq.~\eqref{eq:hydro_final}.
Here, we consider pressure-driven hydrodynamic flow for a small pressure difference across a de Laval nozzle (resulting from small chemical potential or temperature differences), and neglect the electric and magnetic fields.

\marked{
Note that a common approach to incorporate the electric field in the Navier-Stokes equation while retaining its closed form is the gradual-channel approximation~\cite{Dyakonov1993}. Within this approximation, the local electric potential $U$ is related to the local charge carrier density $n$ via an effective capacitance per unit area $C$, such that $n = C U / e$. Taking the gradient, this implies that the electric field is proportional to the carrier density gradient.
For conventional electron fluids, the electric field term in the Navier-Stokes equation can then be absorbed in the gradient of an effective pressure.
However, here we consider a hydrodynamic system of massless Dirac electrons, of which the pressure is related to the energy-dependent mass density [see Eq.~\eqref{eq:PMV}], which is not proportional to the carrier density [see Eqs.~\eqref{eq:D2N}-\eqref{eq:D3M}] such that pressure and charge carrier density gradients cannot be treated on equal footing.
Furthermore, our analytical approach of the effective nozzle equations relies explicitly on momentum conservation [see second equality of Eq.~\eqref{eq:nozzle}], which is violated once the electric field term is retained in Eq.~\eqref{eq:hydro_apx_4}.
On the other hand, the relation $n = C U / e$ shows that the electric field becomes negligible when the effective capacitance is sufficiently large. This is the assumption under which we derive the effective nozzle equations with pressure-driven hydrodynamic flow (see Appendix~\ref{sec:nozzle}).
}

The speed of sound can be easily obtained by linearizing Eqs.~\eqref{eq:hydro_apx_3} and \eqref{eq:hydro_apx_4} around a fluid at rest ($\vecV = 0$) with $E = E_0 + \delta E$, $P = P_0 + \delta P$, $\rttensor{\Pi} = P_0 + \delta P$. We obtain
\begin{align}
	\partial_t \vecS + \nabla (\delta P) &= 0, \\
	\partial_t (\delta E) + v^2 \nabla \cdot \vecS &=0.
\end{align}
The equations can be combined to form a wave equation
\begin{equation}
	\partial^2_t (\delta E) - v_\sound^2 \nabla^2 (\delta E) = 0,
\end{equation}
if one takes into account the definition of the sound velocity as $v_\sound^2 = v^2 \deriv P/\deriv E$.

\section{Dirac electron nozzle}
\label{sec:nozzle}
\begin{figure}
	\centering
	\includegraphics[scale=.3]{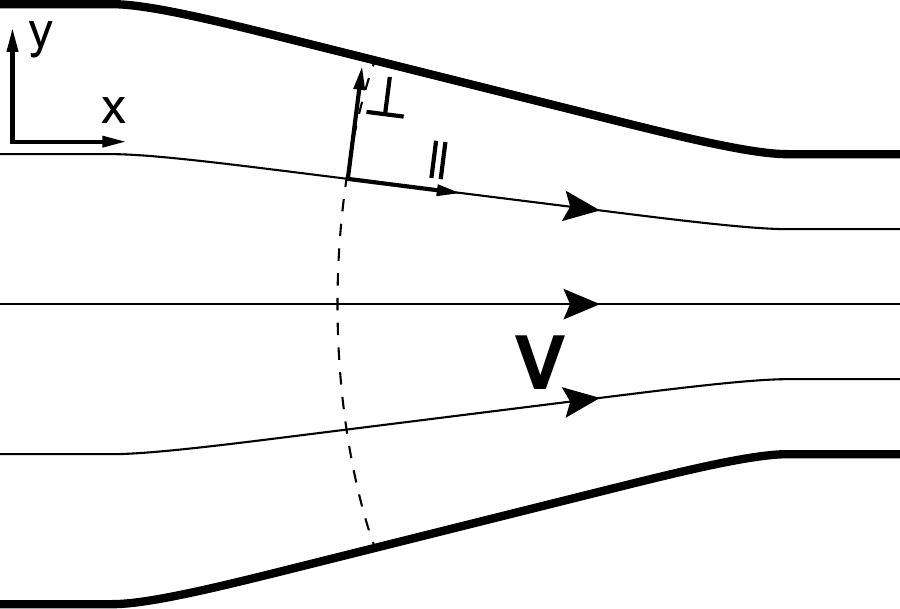}
	\caption{Two-dimensional nozzle geometry with local coordinate system that aligns with the direction of the flow.}
	\label{fig:flow}
\end{figure}
We apply the hydrodynamic equations of Eq.~\eqref{eq:hydro_final_apx} to resolve the velocity profile of a nozzle geometry (see Fig.~\ref{fig:flow}).
We rewrite the velocity $\vecV = V \vecu$ with unit vector $\vecu$ and we can write $\vecu \cdot \nabla \equiv \partial_\parallel$, such that the relations in Eq.~\eqref{eq:hydro_final} become:
\begin{align} \label{eq:hydro_nozzle_1}
	\partial_\parallel (N V) + N V \, \nabla \cdot \vecu &= 0, \\ \label{eq:hydro_nozzle_2}
	\partial_\parallel (M V) + M V \, \nabla \cdot \vecu &= 0, \\ \label{eq:hydro_nozzle_3}
	\partial_\parallel P + M V \partial_\parallel V &= 0, \\ \label{eq:hydro_nozzle_4}
	(\vecu_\perp \cdot \nabla) P - M V^2 \vecu \cdot \partial_\parallel \vecu_\perp &= 0,
\end{align}
with the last line valid for any unit vector $\vecu_\perp \perp \vecu$.
The divergence of the normalized flow vector is related to the increase or decrease of the cross section of the nozzle by
\begin{equation}
	\nabla \cdot \vecu = \frac{\partial_\parallel A}{A},
\end{equation}
assuming laminar flow and thereby ruling out turbulent flow.
Inserting this into Eqs.~\eqref{eq:hydro_nozzle_1} and \eqref{eq:hydro_nozzle_2} and adding Eq.~\eqref{eq:hydro_nozzle_3}, we retrieve the nozzle equations in Eq.~\eqref{eq:nozzle}, where the subscript of the partial derivative, indicating that it acts along the direction of the flow, is omitted. The last equation derived here, Eq.~\eqref{eq:hydro_nozzle_4}, describes how a flow profile makes corners and does not affect the nozzle effect.
Here, we do not explicitly treat the transverse direction(s) of the flow profile and consider the flow to be effectively one-dimensional (along the direction of $\vecu$).

Now we can relate the cross section to the flow speed. Combining the second and third equality of Eq.~\eqref{eq:nozzle}, we obtain:
\begin{equation} \label{eq:nozzle_N_m}
	\frac{\partial N}{N} = \frac{\partial M}{M} = - \frac{\partial(V A)}{V A},
\end{equation}
with the subscript of the partial derivative omitted to simplify the notation.
Combining the first equality of Eq.~\eqref{eq:nozzle} with the expression for the pressure in Eq.~\eqref{eq:macrovalues_eq}, we get:
\begin{equation} \label{eq:mass-velocity}
	\frac{v^2 - V^2}{V^2} \frac{\partial M}{M} + (D - 1) \frac{\partial V}{V} = 0.
\end{equation}
These equations can be combined to obtain the nozzle equations in Eq.~\eqref{eq:nozzle_sol}.

To derive the nozzle equations in terms of temperature and chemical potential [Eq.~\eqref{eq:nozzle_3}], some additional manipulations are required. Let us separate the velocity dependence,
\begin{align}
\begin{split}
	N &= N_{0}[1-(V/v)^{2}]^{-(D+1)/2}, \\
	M &=  M_{0}[1-(V/v)^{2}]^{-(D+3)/2},
\end{split}
\end{align}
where we define $M_{0}=\left.M\right|_{V=0}$ and $N_{0}=\left.N\right|_{V=0}$. These definitions can be used to rewrite Eqs.~\eqref{eq:nozzle_N_m}-\eqref{eq:mass-velocity} as follows:
\begin{align} \label{eq:nozzle_1_apx}
	\frac{\partial V}{V} &= -\frac{1-(V/v)^{2}}{1-D(V/v)^{2}} \frac{\partial A}{A}, \\ \label{eq:N_0}
    \frac{\partial N_0}{N_0} &= \frac{2D(V/v)^{2}}{1-D(V/v)^{2}}\frac{\partial A}{A}, \\ \label{eq:m_0}
	\frac{\partial M_0}{M_0} &= \frac{2(D+1)(V/v)^{2}}{1-D(V/v)^{2}}\frac{\partial A}{A},
\end{align}
Let us parametrize $N_0$ and $M_0$ as functions of the temperature $T$ and the ratio of chemical potential and temperature $\mu/T$:
\begin{align} \label{eq:CopyMNDexplicit}
	N_0 &\propto T^D F_D^-(\mu/T),
	&
	M_0 &\propto T^{D+1} F_{D+1}^+(\mu/T).
\end{align}
Then Eqs.~\eqref{eq:N_0} and~\eqref{eq:m_0} transform into
\begin{align}
	D\frac{\partial T}{T} + \frac{\partial F_D^-(\mu/T)}{F_D^-(\mu/T)} &= 2D \frac{(V/v)^2}{1 - D(V/v)^2} \frac{\partial A}{A},
\end{align}
\begin{align}
\begin{split}
	&(D+1) \frac{\partial T}{T} + \frac{\partial F_{D+1}^{+}(\mu/T)}{F_{D+1}^{+}(\mu/T)} \\
    &\qquad = 2(D+1) \frac{(V/v)^2}{1 - D(V/v)^2} \frac{\partial A}{A},
\end{split}
\end{align}
having substituted $\partial V/V$ through Eq.~\eqref{eq:nozzle_1_apx}. Solving these equations for the partial derivative acting on the argument of the gamma functions, we obtain $\partial(\mu/T) = 0$, which then yields
\begin{align}
	\frac{\partial T}{T} &= \frac{\partial \mu}{\mu},
	&
	\frac{\partial T}{T} &= \frac{2(V/v)^2}{1 - D(V/v)^2}\frac{\partial A}{A}.
	\label{eq:nozzle_2_apx}
\end{align}

The solution of Eq.~\eqref{eq:nozzle_1_apx} is
\begin{equation} \label{eq:CA_apx}
	V^2 (v^2 - V^2)^{D-1} A^2 = C_A^2 v^{2D} = \textnormal{const.} ,
\end{equation}
with $C_A$ an integration constant that fixes the relation between the cross section of the nozzle and the flow speed.
Note that there is an upper limit for $C_A/A$, namely,
\begin{align}
\begin{split}
	&0 \leq \frac{C_A}{A} \leq \kappa, \\
	&\kappa \equiv \frac{(D-1)^{(D-1)/2}}{D^{D/2}} = \begin{cases} 1/2 & (D = 2) \\ 2/(3\sqrt3) & (D = 3) \end{cases},
\end{split}
\end{align}
The solutions are presented in Fig.~\ref{fig:CA_CP_CX}a.
The solution of Eq.~\eqref{eq:nozzle_2_apx} is given by
\begin{align}
\begin{split}
	\frac{\mu}{v^2 - V^2} &= \frac{C_\mu}{v^2} = \textnormal{const.},
	\\
	\frac{T}{v^2 - V^2} &= \frac{C_T}{v^2} = \textnormal{const.},
	\label{eq:CmuT_apx}
\end{split}
\end{align}
with $C_{\mu,T}$ the chemical potential and temperature for vanishing flow speed (see Fig.~\ref{fig:CA_CP_CX}c). The formulae for $\mu$ and $T$ have the same form. We can obtain the dependence on the cross section $A$ by substituting Eq.~\eqref{eq:CmuT_apx} into Eq.~\eqref{eq:CA_apx}, resulting in
\begin{equation} \label{eq:CX_apx}
	\frac{\mu}{C_\mu} \lef 1 - \frac{\mu}{C_\mu} \rig^{D-1} = \frac{C_A^2}{A^2}.
\end{equation}
and an identical equation for $T/C_T$. The hydrodynamic equations cannot independently match the chemical potential and temperature at the entrance and exit of the nozzle with the values in the leads.
For example, in the motionless case, i.e., $V=0$, the chemical potential $\mu(\vecr)$ and temperature $T(\vecr)$ can be coordinate-dependent while the pressure $P = M (v^2 - V^2)/(D+1)$ [equal to $M_0 v^2/(D+1)$ for $V=0$] remains constant, so that the flow gradient is zero and the hydrodynamic Navier-Stokes equation does not induce any flow.
Thus, we should always match the pressure of the leads and cannot match both $\mu$ and $T$.

\subsection{Pressure-driven flow}
\label{subsec:pressure}
\begin{figure*}[tb]
	\centering
	\includegraphics[height=.2\linewidth]{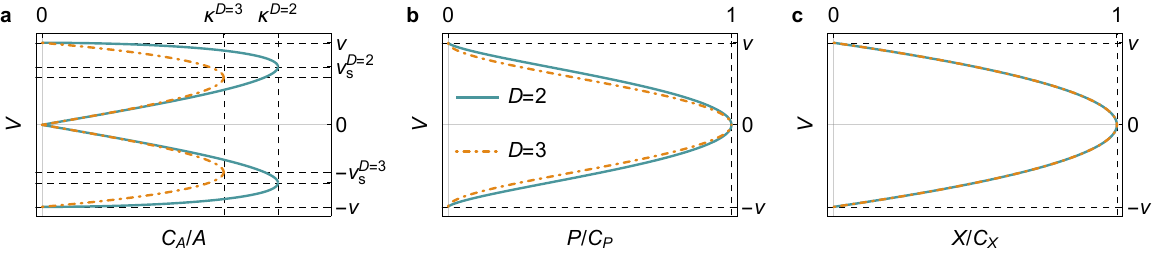}
	\caption{
		(\textbf{a})-(\textbf{c}) The relation between flow speed and (\textbf{a}) cross section, (\textbf{b}) pressure, and (\textbf{c}) chemical potential or temperature, according to Eqs.~\eqref{eq:CA_apx}, \eqref{eq:CP_CA_apx} and \eqref{eq:CmuT_apx}, respectively, for two-dimensional (blue) and three-dimensional (yellow, dash-dotted) nozzles.
	}
	\label{fig:CA_CP_CX}
\end{figure*}

An explicit expression for the pressure in terms of temperature, chemical potential and flow speed can be obtained from Eq.~\eqref{eq:mass} and the relation for the pressure in Eq.~\eqref{eq:macrovalues_eq}, resulting in:
\begin{equation} \label{eq:pressure_apx}
\begin{split}
	P &= \frac{S^D}{(2 \pi)^D} \frac{\Gamma(D + 1)}{D} \frac{v T^{D + 1}}{(v^2 - V^2)^{(D + 1)/2}} F_{D + 1}^+(\mu/T) \\
	&= C_P \lef 1 - \frac{V^2}{v^2} \rig^{(D+1)/2},
\end{split}
\end{equation}
where the last line is obtained with Eq.~\eqref{eq:CmuT_apx} and $C_P$ is given by:
\begin{equation} \label{eq:CP_apx}
	\begin{split}
		C_P &= \frac{S^D}{(2 \pi)^D} \frac{\Gamma(D + 1)}{D} \frac{C_T^{D + 1}}{v^D} F_{D + 1}^+(C_\mu/C_T) \\
		&\approx \frac{S^D}{(2 \pi)^D} \frac{\Gamma(D + 1)}{D} \frac{1}{v^D} \\
        &\hphantom{\approx} \times \begin{cases} 3 \zeta_3 C_T^3 / 2 & (C_\mu \ll C_T, D = 2) \\ 2(1 - 2^{-D})\zeta_{D+1} C_T^{D + 1} & (C_\mu \ll C_T, D \neq 2) \\ |C_\mu|^{D + 1} / (D + 1)! & (C_\mu \gg C_T) \end{cases},
	\end{split}
\end{equation}
which can be interpreted as the pressure for vanishing flow speed, analogous to $C_\mu$ and $C_T$ being the chemical potential and temperature for vanishing flow speed, respectively.
The relation is presented for $D = 2$ and $D = 3$ in Fig.~\ref{fig:CA_CP_CX}b.
The pressure can also be related to the cross section $A$ by combining Eq.~\eqref{eq:CA_apx} with Eq.~\eqref{eq:pressure_apx}:
\begin{equation} \label{eq:CP_CA_apx}
\begin{split}
	\lef \frac{C_A}{A} \rig^2
	&= \frac{V^2}{v^2} \lef 1 - \frac{V^2}{v^2} \rig^{D - 1} \\
	&= \left[ 1 - \lef \frac{P}{C_P} \rig^{-2/(D+1)} \right]
	\lef \frac{P}{C_P} \rig^{- 2\frac{D - 1}{D + 1}},
\end{split}
\end{equation}
as presented in Eq.~\eqref{eq:CP}.

\begin{figure*}[tb]
	\centering
	\includegraphics[height=0.2\linewidth]{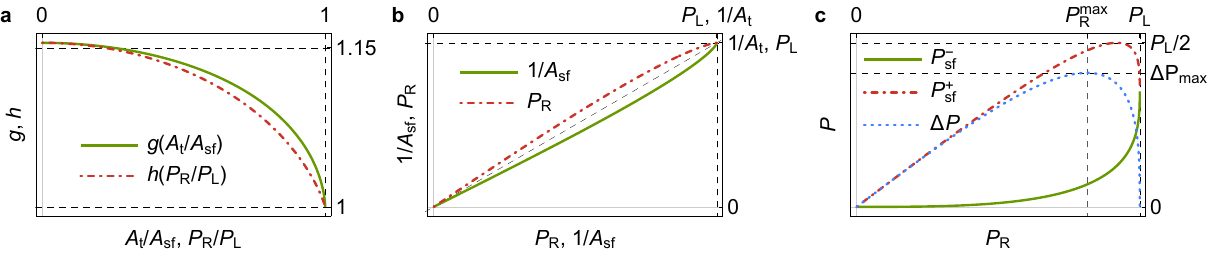}
	\caption{
		(\textbf{a}) Relations $g$ and $h$ of Eq.~\eqref{eq:A_sw} as a function of the ratio of cross sections of the throat and the shock front and of the ratio of pressures in the leads, respectively.
		(\textbf{b}) The relation between the cross section of the shock front relative to the throat and the pressure in the leads.
		(\textbf{c}) The pressure to the left and right of the shock front as a function of the pressure in the right lead, as well as the size of the pressure jump.
	}
	\label{fig:A_sw}
\end{figure*}

The momentum $S$ is conserved throughout the nozzle identical for any position along the nozzle and is given by:
\begin{equation} \label{eq:heatflow_apx}
\begin{split}
	S &= M V A \\
    &= \frac{S^D}{(2 \pi)^D} \frac{D + 1}{D} \Gamma(D + 1) \frac{v T^{D+1} V A}{(v^2 - V^2)^{(D+3)/2}} F_{D+1}^+(\mu/T) \\
    &= (D + 1) C_P C_A / v,
\end{split}
\end{equation}
where the last equality is obtained using Eqs.~\eqref{eq:CA_apx} and \eqref{eq:pressure_apx}. The same conservation law applies to the energy flow $v^2 S$.
Thus, we obtain the following relation:
\begin{equation} \label{eq:heat_flow_cond}
	C_P^\Le C_A^\Le = C_P^\Ri C_A^\Ri \quad \Longrightarrow \quad P_\Le C_A^\Le = P_\Ri C_A^\Ri.
\end{equation}
with pressures $P_\Le$ and $P_\Ri$ for the left and right leads, respectively, and the superscript denoting whether the constant belongs to the solution to the left or to the right of the shock front (see discussion of supersonic flow profile in Sec.~\ref{sec:pressure-driven-flow}).
Combining this relation with Eq.~\eqref{eq:CP_CA_apx} and the expression for $C_A^\Le$ in the case of a supersonic flow profile from left to right, i.e., $C_A^\Le = \kappa A_\throat$ (with minimal cross section $A_\throat$ at the throat of the nozzle) we obtain the following relations between the cross section and the pressure of the nozzle, to the left and right of the shock front, respectively:
\begin{align}
    \label{eq:pressure-left}
    \left(\frac{\kappa A_\throat}{A}\right)^{2} &=
	\left[1 - \left(\frac{P}{P_\Le}\right)^{-2\frac{1}{D+1}}\right]
	\left(\frac{P}{P_\Le}\right)^{- 2\frac{D - 1}{D + 1}}, \\
    \label{eq:pressure-right}
    \left(\frac{P_\Le}{P_\Ri}\frac{\kappa A_\throat}{A}\right)^{2} &=
	\left[1 - \left(\frac{P}{P_\Ri}\right)^{-2\frac{1}{D+1}}\right]
	\left(\frac{P}{P_\Ri}\right)^{- 2\frac{D - 1}{D + 1}}.
\end{align}
To determine the position of the shock front $x_\sw$ and its cross section $A_\sw$, we infinitesimally integrate Eq.~\eqref{eq:nozzle} across the discontinuity.
The integration of the second equality gives
\begin{align}
	\Delta(MV) &= 0,
\end{align}
which tells us that $MV$ is conserved across the discontinuity.
That property can be used in the integration of the first equality, yielding
\begin{align}
	\Delta P + MV \Delta V &= 0,
	&
	\Delta(P + MV^2) &= 0.
\end{align}
Inserting the relation for the pressure of Eq.~\eqref{eq:macrovalues_eq}, we get:
\begin{equation} \label{eq:jump}
	\Delta \lef v^2 / V + DV \rig = 0.
\end{equation}
We proceed to solve this equation for a Dirac system in two spatial dimensions ($D = 2$).
Using Eq.~\eqref{eq:CP_CA_apx}, we can express the flow speeds just in front and beyond the jump in terms of the cross section of the throat and the shock front, and the pressure in the leads:
\begin{align} \label{eq:V_sw}
\begin{split}
	V_\sw^- &= v_\sound \lef 1 + \sqrt{1 - \lef \frac{A_\throat}{A_\sw} \rig^2} \rig^{1/2}, \\
	V_\sw^+ &= v_\sound \lef 1 - \sqrt{1 - \lef \frac{P_\Le}{P_\Ri} \frac{A_\throat}{A_\sw} \rig^2} \rig^{1/2},
\end{split}
\end{align}
\begin{align}
\label{eq:P_sw}
\begin{split}
	P_\sw^- &= P_\ast \lef 1 - \sqrt{1 - \lef \frac{A_\throat}{A_\sw} \rig^2} \rig^{3/2}, \\
	P_\sw^+ &= P_\ast \frac{P_\Ri}{P_\Le} \lef 1 + \sqrt{1 - \lef \frac{P_\Le}{P_\Ri} \frac{A_\throat}{A_\sw} \rig^2} \rig^{3/2},
\end{split}
\end{align}
with critical pressure $P_\ast = P_\Le / 2^{3/2}$ and $Z_\sw^\pm \equiv \lim_{\delta \rightarrow 0} Z(x_\sw \pm \delta)$ with $Z = P, V$.
The flow speed (pressure) is supersonic (subcritical) on the left side of the shock front and subsonic (supercritical) on the right. Inserting the flow speeds into Eq.~\eqref{eq:jump}, we can relate the ratio of pressures in the leads to the ratio of cross sections for the throat and the shock front, leading to the following relation:
\begin{widetext}
\begin{equation} \label{eq:sol_Asw}
	\frac{P_\Ri}{P_\Le} = \frac{A_\throat}{A_\sw} \frac{\sqrt{(A_\throat/A_\sw)^2 \lef 3 + 2\sqrt{1 - (A_\throat/A_\sw)^2}\rig - 2 \lef 1 + \sqrt{1-(A_\throat/A_\sw)^2} \rig}}{4 (A_\throat/A_\sw)^2 - 3}.
\end{equation}
\end{widetext}
The last factor in the equality is approximately equal to one such that $P_\Ri / P_\Le \approx A_\throat/A_\sw$, with a deviation of at most $\sim$15\% (see Fig.~\ref{fig:A_sw}b).
We can write
\begin{equation} \label{eq:A_sw}
	g\lef \frac{A_\throat}{A_\sw} \rig = \frac{A_\sw}{A_\throat} \frac{P_\Ri}{P_\Le} = h\lef \frac{P_\Ri}{P_\Le} \rig,
\end{equation}
with $g$ written explicitly in Eq.~\eqref{eq:sol_Asw} and $h$ obtained by solving the equation for $A_\sw / A_\throat$ instead. These functions are shown in Fig.~\ref{fig:A_sw}a.
The pressure at the shock front has a jump $\Delta P$, which can be obtained explicitly from Eqs.~\eqref{eq:P_sw} and \eqref{eq:sol_Asw}, and is shown in Fig.~\ref{fig:A_sw}c.
If we fix $P_\Le$ and vary $P_\Ri$, the maximal pressure jump $\Delta P_{\max}$ is given by the maximum of the function $\Delta P/P_\Le$, which leads to the following values for the cross section of the shock front, the pressure in the right lead, and the size of the pressure jump:
\begin{align}
    \begin{aligned}
    	A_\sw^{\max} &\approx 1.34 \, A_\throat, \\
        \Delta P_{\max} &\approx 0.41 \, P_\Le,
    \end{aligned} &&
	\begin{aligned}
	    P_\Ri^{\max} &\approx 0.81 \, P_\Le, \\
        &
	\end{aligned}
\end{align}
which are also indicated in Fig.~\ref{fig:A_sw}c.
An example of a supersonic flow profile with finite pressure difference and discontinuity in pressure and flow velocity at the shock front position is presented in Fig.~\ref{fig:distribution}.

\subsection{Chemical potential and temperature}
\label{subsec:mu_and_T}
We have seen that, to obtain a flow from left to right, we need $P_\Le > P_\Ri$. In terms of the temperature and chemical potential, we can see from Eq.~\eqref{eq:CP_apx} that this translates to the following condition:
\begin{equation}
	\lef \frac{T_\Le}{T_\Ri} \rig^{D + 1} \frac{F_{D + 1}^+(\mu_\Le / T_\Le)}{F_{D + 1}^+(\mu_\Ri/T_\Ri)} > 1.
\end{equation}
An equal pressure is obtained when the left-hand side is equal to one, as required for a subsonic flow profile. In the high- and low temperature regimes of the leads, $\mu_{\Le, \Ri} \ll T_{\Le, \Ri}$ and $\mu_{\Le, \Ri} \gg T_{\Le, \Ri}$, respectively, the condition simplifies to $T_\Le > T_\Ri$ and $\mu_\Le > \mu_\Ri$,
making use of the expansion in Eq.~\eqref{eq:F_exp_1}.
As expected, the temperature (chemical potential) gradient determines the direction of the flow in the regime where the temperature (chemical potential) dominates, with the flow going from high to low temperature (chemical potential).

To resolve the constants for the chemical potential and temperature profiles that correspond to the solution of the nozzle equations, $C_\mu$ and $C_T$, we need another relation in addition to Eq.~\eqref{eq:heat_flow_cond}, which originates from momentum (or equivalently, energy flow) conservation.
Recall that the integration constants for temperature and chemical potential cannot independently be matched to the corresponding values in the leads because the Navier-Stokes equation only ensures that the pressure matches. At the lead where the flow originates, the integration constants for temperature and chemical potential inherit the values from the lead, whereas, at the exit side, they follow from the conservation of current and momentum along the nozzle.

In addition to momentum conservation, we make use of the conservation of particle current $I$, which is equal to:
\begin{equation} \label{eq:current_apx}
\begin{split}
	I &= N V A \\
    &= \frac{S^D}{(2 \pi)^D} \Gamma(D) \frac{v T^D V A}{(v^2 - V^2)^{(D + 1)/2}} F_D^-(\mu/T) \\
	&= \frac{S^D}{(2 \pi)^D}  \frac{\Gamma(D)}{v^{D-1}}  C_A C_T^D F_D^-(C_\mu/C_T),
\end{split}
\end{equation}
which follows from Eqs.~\eqref{eq:N_apx}, \eqref{eq:CA_apx} and \eqref{eq:CmuT_apx}.
Analogously as for the momentum in Eq.~\eqref{eq:heatflow_apx}, we obtain the last equality by considering an infinitely wide lead with $V = 0$.

Let us now consider a flow that goes from left to right, and work out the flow profile and the corresponding profiles for the chemical potential and temperature. In this case, we have $C_\mu^\Le = \mu_\Le$, $C_T^\Le = T_\Le$ and the following relations hold:
\begin{widetext}
\begin{align} \label{eq:sol_matching}
	\begin{cases} C_A^\Le T_\Le^D F_D^-(\mu_\Le/T_\Le) = C_A^\Ri (C_T^\Ri)^D F_D^-(C_\mu^\Ri/C_T^\Ri) \\
	C_A^\Le T_\Le^{D + 1} F_{D + 1}^+(\mu_\Le/T_\Le) = C_A^\Ri (C_T^\Ri)^{D + 1} F_{D + 1}^+(C_\mu^\Ri/C_T^\Ri) \\
	(C_T^\Ri)^{D + 1} F_{D + 1}^+(C_\mu^\Ri/C_T^\Ri) = (T_\Ri)^{D + 1} F_{D + 1}^+(\mu_\Ri/T_\Ri)
    \end{cases},
\end{align}
which follow from matching the current and the momentum in both leads and from matching the pressure in the right lead, respectively, making use of Eqs.~\eqref{eq:CP_apx}, \eqref{eq:heatflow_apx} and \eqref{eq:current_apx}.
We separate the cases of subsonic and supersonic flow:
\begin{align} \label{eq:sol_subsonic}
	\textnormal{subsonic: } C_A^\Le &= C_A^\Ri = \frac{(2 \pi)^D}{S^D} \frac{1}{\Gamma(D)} v^{D - 1} \frac{I}{T_\Le^D F_D^-(\mu_\Le/T_\Le)}, \quad I < I_{\max}, \\ \label{eq:sol_supersonic}
	\textnormal{supersonic: } C_A^\Le &= \kappa A_\throat, \quad C_A^\Ri = \kappa A_\throat \frac{T_\Le^{D + 1}}{T_\Ri^{D + 1}} \frac{F_{D + 1}^+(\mu_\Le / T_\Le)}{F_{D + 1}^+(\mu_\Ri / T_\Ri)}, \quad I = I_{\max},
\end{align}
\begin{align} \label{eq:I_max}
	\begin{split}
		I_{\max} &= \frac{S^D}{(2 \pi)^D} \Gamma(D) \frac{1}{v^{D - 1}} \kappa A_\throat T_\Le^D F_D^-(\mu_\Le/T_\Le) \\
		&\approx \frac{S^D}{(2 \pi)^D} \Gamma(D) \frac{1}{v^{D - 1}} \kappa A_\throat \times \begin{cases} 2 \ln 2 \, T_\Le^2 (\mu_\Le / T_\Le) & (\mu_\Le \ll T_\Le, D = 2) \\ 2 (1 - 2^{2 - D}) \zeta_{D - 1} T_\Le^D (\mu_\Le / T_\Le) & (\mu_\Le \ll T_\Le, D \neq 2) \\ (\sign \, \mu_\Le) | \mu_\Le |^D / D! & (\mu_\Le \gg T_\Le) \end{cases}.
	\end{split}
\end{align}
\end{widetext}
These relations are sufficient to extract the values of $C_{\mu, T}$ and reconstruct the profiles in the nozzle via Eq.~\eqref{eq:CX_apx}.

We proceed here by explicitly considering the case of $D = 2$. The solution of Eq.~\eqref{eq:CX_apx} is then given by:
\begin{equation}
	\frac{\mu}{C_\mu} = \frac{1}{2} \lef 1 \pm \sqrt{1 - \lef \frac{2 C_A}{A} \rig^2} \rig.
\end{equation}
and an identical solution for $T/C_T$. In the subsonic regime, we get the following profile in the nozzle

\begin{align} \label{eq:CmuT_left_apx}
\begin{split}
	\mu &= \frac{\mu_\Le}{2} \lef 1 + \sqrt{1 - \lef \frac{2 C_A^\Le}{A} \rig^2} \rig,
	\\
	T &= \frac{T_\Le}{2} \lef 1 + \sqrt{1 - \lef \frac{2 C_A^\Le}{A} \rig^2} \rig,
\end{split}
\end{align}
with $C_A^\Le$ related to the current via Eq.~\eqref{eq:sol_subsonic}.
In the supersonic regime, the profile to the left of the shock front is given by:
\begin{align} \label{eq:CmuT_left_super_apx}
\begin{split}
    \mu &= \frac{\mu_\Le}{2} \lef 1 \pm \sqrt{1 - \lef \frac{A_\throat}{A} \rig^2} \rig,
	\\
	T &= \frac{T_\Le}{2} \lef 1 \pm \sqrt{1 - \lef \frac{A_\throat}{A} \rig^2} \rig,
\end{split}
\end{align}
where the $+$ ($-$) sign corresponds to the solution to the left (right) of the throat.
Past the shock front, we get:
\begin{align} \label{eq:CmuT_right_super_apx}
\begin{split}
	\mu &= \frac{C_\mu^\Ri}{2} \lef 1 + \sqrt{1 - \lef \frac{2 C_A^\Ri}{A} \rig^2} \rig,
	\\
	T &= \frac{C_T^\Ri}{2} \lef 1 + \sqrt{1 - \lef \frac{2 C_A^\Ri}{A} \rig^{2}} \rig,
\end{split}
\end{align}
To solve for the values of $C_\mu^\Ri$ and $C_\mu^\Ri$ and the resulting profile beyond the shock front, we can use the two independent equations that remain from Eq.~\eqref{eq:sol_matching}:
\begin{align}
    \begin{cases}
	   (C_T^\Ri)^2 F_2^-(C_\mu^\Ri/C_T^\Ri) =  \frac{T_\Ri^3}{T_\Le} \frac{F_3^+(\mu_\Ri/T_\Ri) F_2^-(\mu_\Le/T_\Le)}{F_3^+(\mu_\Le/T_\Le)} \\
	   (C_T^\Ri)^3 F_3^+(C_\mu^\Ri/C_T^\Ri) = T_\Ri^3 F_3^+(\mu_\Ri/T_\Ri)
    \end{cases},
\end{align}
making use of the relation between $C_A^\Ri$ and $C_A^\Le$ in the supersonic regime, given by Eq.~\eqref{eq:sol_supersonic}.
Let us first consider the limit regime $T_{\Le,\Ri} \gg \mu_{\Le,\Ri}$. Then we have:
\begin{align} \label{eq:CmuT_sol_highT_apx}
\begin{split}
    &\begin{cases}
		\displaystyle
		(C_T^\Ri)^2 F_2^-(C_\mu^\Ri / C_T^\Ri) = 2 \ln 2 \, T_\Ri^3 \mu_\Le / T_\Le^2
		\\
		(C_T^\Ri)^3 F_3^+(C_\mu^\Ri / C_T^\Ri) = 3\zeta_3 T_\Ri^3 / 2
	\end{cases}
	\\ \Rightarrow &
	\begin{cases}
		\displaystyle
		C_\mu^\Ri = T_\Ri^2 \mu_\Le / T_\Le^2 \ll T_\Ri
		\\
		C_T^\Ri = T_\Ri
	\end{cases}.
\end{split}
\end{align}
We see that the exit temperature matches the value in the right lead, whereas the chemical potential does not (see Fig.~\ref{fig:CX_TR_muR}a).
For the opposite limit regime, with $T_{\Le,\Ri} \ll \mu_{\Le,\Ri}$, we obtain:
\begin{equation} \label{eq:CmuT_sol_lowT_apx}
	\begin{cases}
		\displaystyle
		(C_T^\Ri)^2 F_2^-(C_\mu^\Ri/C_T^\Ri) = \mu_\Ri^3 / (2 \mu_\Le)
		\\
		(C_T^\Ri)^3 F_3^+(C_\mu^\Ri/C_T^\Ri) = \mu_\Ri^3 / 6
	\end{cases}.
\end{equation}
In this case, a small exit temperature, $C_T^\Ri \ll C_\mu^\Ri$, is not guaranteed.
In the limit of very small chemical potential difference, $\mu_\Le - \mu_\Ri \ll \mu_\Le$, we can expand the left-hand side of both equations, using the expansions in Eqs.~\eqref{eq:F_exp_1} and \eqref{eq:F_exp_2}, yielding:
\begin{align}
\begin{split}
	& \begin{cases}
		(C_\mu^\Ri)^2 + \pi^2 (C_T^\Ri)^2 / 3 = \mu_\Ri^3 / \mu_\Le \\
		(C_\mu^\Ri)^3 + \pi^2 C_\mu^\Ri (C_T^\Ri)^2 = \mu_\Ri^3
	\end{cases}
	\\ \Rightarrow &
	\begin{cases}
		C_\mu^\Ri = 2 \mu_\Ri- \mu_\Le \\
		C_T^\Ri = \sqrt{3 (\mu_\Le - \mu_\Ri) \mu_\Ri} / \pi
	\end{cases}.
\end{split}
\label{eq:CmTasymptoteReqL}
\end{align}
In the opposite limit, $\mu_\Le \gg \mu_\Ri$, we can consider the following expansion and corresponding solution:
\begin{align}
\begin{split}
	& \begin{cases}
		2\ln2 \, C_\mu^\Ri C_T^\Ri = \mu_\Ri^3 / (2 \mu_\Le) \\
		3 \zeta_3 (C_T^\Ri)^3 + 2 \ln2 \, (C_\mu^\Ri)^2 C_T^\Ri = \mu_\Ri^3 / 3
	\end{cases}
	\\ \Rightarrow &
	\begin{cases}
		C_\mu^\Ri = \sqrt[3]{9 \zeta_3} \mu_\Ri^2 / (4 \ln 2 \, \mu_\Le) \\
		C_T^\Ri = \mu_\Ri / \sqrt[3]{9\zeta_{3}}.
	\end{cases}.
\end{split}
\label{eq:CmTasymptoteRllL}
\end{align}
The general solution of Eq.~\eqref{eq:CmuT_sol_lowT_apx} for the low entrance-temperature regime is shown in Fig.~\ref{fig:CX_TR_muR}b, together with the asymptotes obtained in Eqs.~\eqref{eq:CmTasymptoteReqL} and~\eqref{eq:CmTasymptoteRllL}.
Note that there is significant hydrodynamic heating in general, with the exit temperature being proportional to the chemical potential in the entrance lead, which is considered to be much larger than the temperature in the leads.
Only when $\mu_\Ri = \mu_\Le$ does the chemical potential at the nozzle exit match with the right lead.

\begin{figure*}[tb]
	\centering
	\includegraphics[width=.9\linewidth]{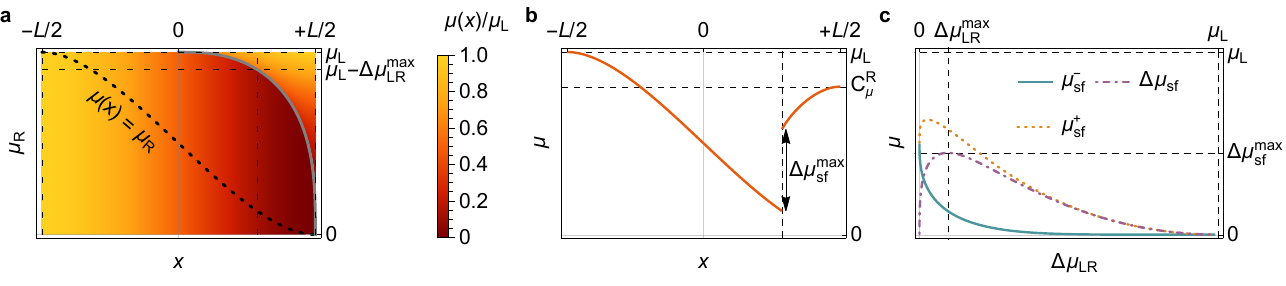}
	\caption{
		(\textbf{a}) The local chemical potential as a function of the nozzle coordinate (ranging between $-L/2$ and $+L/2$, for a two-dimensional nozzle with infinitely wide leads, as considered in Fig.~\ref{fig:distribution}) and the chemical potential of the right lead, in the limit $T_{\Le, \Ri} \ll \mu_\Le$.
        The gray line indicates the position of the shock front.
		(\textbf{b}) The spatial profile of the chemical potential for a chemical potential difference that yields the maximal jump of the chemical potential at the shock front, as indicated by a horizontal dashed line in \textbf{a}.
         (\textbf{c}) The values of the chemical potential at the shock front position, approached from the left ($\mu_\sw^-$) and from the right ($\mu_\sw^+$), as well as the size of the chemical potential jump $\Delta \mu_\mathrm{sf} = \mu_\sw^+ - \mu_\sw^-$, as a function of the chemical potential difference over the de Laval nozzle, with the chemical potential difference that yields the maximal jump, $\Delta\mu_\mathrm{LR}^{\max}$ ($\approx 0.095 \, \mu_\Le$), indicated.
	}
	\label{fig:voltage-dL}
\end{figure*}

\subsection{Voltage characteristics}
\label{subsec:voltage}
Based on the results of the previous section, we can derive the voltage characteristics of a de Laval nozzle with hydrodynamic Dirac electrons. Here, we consider the low-temperature regime ($T_{\Le,\Ri} \ll \mu_{\Le,\Ri}$) of a nozzle with infinitely wide leads and, without loss of generality, consider a flow of electrons from left to right ($\mu_\Le \geq \mu_\Ri \geq 0$), keeping $\mu_\Le$ fixed and $\mu_\Ri$ variable.
Note that we ignore the heating effect across the nozzle here.
The voltage across the de Laval nozzle, denoted as $U_\mathrm{dL}$, is then related to the chemical potential difference between the leads as follows:
\begin{equation}
	U_\mathrm{dL} = \Delta\mu_{\Le \Ri}/(-e) = (\mu_\Le - \mu_\Ri)/(-e).
\end{equation}
When momentum relaxation and viscosity can be neglected, any finite voltage difference necessarily induces a pressure difference across the nozzle and consequently a supersonic flow profile, with current pinned to its maximum (with flow speed reaching the speed of sound at the throat), given by Eq.~\eqref{eq:I_max}. We can use the relations of Eq.~\eqref{eq:sol_supersonic} to obtain the flow profile solutions as a function of the chemical potentials in the leads:
\begin{align} \label{eq:voltage-characteristics-current}
\begin{split}
	C_A^\Ri &= C_A^\Le \lef \frac{\mu_\Le}{\mu_\Ri} \rig^{D + 1} = \lef \frac{\mu_\Le}{\mu_\Ri} \rig^{D + 1} \kappa A_\throat, \\
	I &= I_{\max} = \frac{S^D}{(2 \pi)^D} \frac{\Gamma(D)}{\Gamma(D + 1)} \frac{1}{v^{D - 1}} \kappa A_\throat \mu_\Le^D.
\end{split}
\end{align}
The solution constants for temperature and chemical potential follow from Eq.~\eqref{eq:sol_matching}. In the case of $D = 2$, we can use the solution of Eq.~\eqref{eq:CmuT_sol_lowT_apx} and obtain the profile of the chemical potential $\mu(x)$ at any position along the nozzle $x$ for different chemical potential differences (see Figs.~\ref{fig:voltage-dL}a,b).

Similarly to a discontinuity of flow speed and pressure at the shock front position, there is a discontinuity of the chemical potential. The chemical potential jump $\Delta \mu_\mathrm{sf}$ at the shock front can be obtained by combining Eqs.~\eqref{eq:CP_apx}, \eqref{eq:sol_Asw}, \eqref{eq:sol_supersonic}, and \eqref{eq:CmuT_left_super_apx}-\eqref{eq:CmuT_right_super_apx}, yielding
\begin{align}
	\begin{split}
		\Delta \mu_\mathrm{sf} &= \mu_\sw^+ - \mu_\sw^-, \\
		\mu_\sw^- &= \frac{\mu_\Le}{2} \lef 1 - \sqrt{1 - \lef \frac{\mu_\Ri^3}{\mu_\Le^3} \frac{1}{h(\mu_\Ri^3/\mu_\Le^3)} \rig^2} \rig, \\
		\mu_\sw^+ &= \frac{C_\mu^\Ri}{2} \lef 1 + \sqrt{1 - \lef \frac{1}{h(\mu_\Ri^3/\mu_\Le^3)} \rig^2} \rig.
	\end{split}
\end{align}
The result is shown as a function of $\Delta\mu_\mathrm{LR}$ in Fig.~\ref{fig:voltage-dL}c. The chemical potential difference that induces the largest discontinuity of the chemical potential at the shock front, denoted by $\Delta\mu_\mathrm{LR}^{\max}$, can be extracted from the numerical solution for $C_\mu^\Ri$ of Eq.~\eqref{eq:CmuT_sol_lowT_apx}, which leads to $\Delta\mu_\mathrm{LR}^{\max} \approx 0.095 \, \mu_\Le$ with a drop of $\Delta \mu_\mathrm{sf}^{\max} \approx 0.45 \, \mu_\Le$. For this chemical potential difference, the shock front appears to the right of the throat where the cross section of the nozzle is approximately equal to $1.49 \, A_\throat$, as can be obtained from the following relation for the cross section of the shock front as a function of the chemical potential in the leads:
\begin{equation}
	A_\sw = A_\throat \frac{\mu_\Le^3}{\mu_\Ri^3} h(\mu_\Ri^3/\mu_\Le^3).
\end{equation}
This cross section is slightly larger than the one at which the maximal pressure jump occurs (with cross section approximately equal to $1.34 \, A_\throat$, as obtained in Appendix~\ref{subsec:pressure}).

With the local chemical potential profile resolved, we can consider the local voltage difference with respect to the left lead $U_\Le(x) = [\mu_\Le - \mu(x)]/(-e)$.
Similarly, we can define a local differential resistance $\mathrm{d}U_\Le(x)/\mathrm{d} I$ and consider its spatial derivative $\partial_x \mathrm{d}U_\Le(x)/\mathrm{d} I$.
We will evaluate these quantities in the subsection below for a more realistic nozzle setup with Ohmic leads that have a finite width.

\subsection{Nozzle with Ohmic \marked{sections}}
\label{subsec:nozzle-Ohmic}
\begin{figure}[tb]
	\centering
	\includegraphics[width=\linewidth]{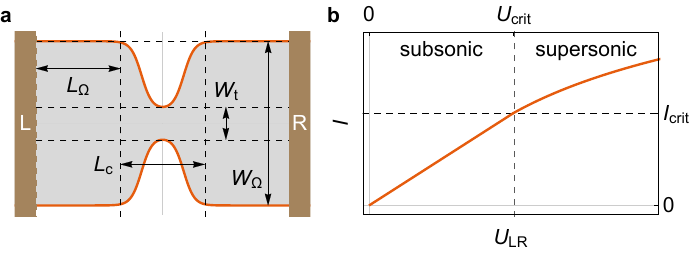}
	\caption{
		(\textbf{a}) A schematic of a de Laval nozzle \marked{consisting of a constriction region in between two Ohmic sections with constant width}, with relevant length scales indicated.
		(\textbf{b}) The current-voltage relation of a de Laval nozzle with Ohmic \marked{sections}, according to the lumped-element circuit model (see Appendix~\ref{subsec:nozzle-Ohmic} for details). The relation is Ohmic up to a critical bias voltage $U_\mathrm{crit}$, where supersonic flow sets in and a further increase in voltage modifies the current-voltage relation from a purely Ohmic one to one that accounts for the voltage being split over the Ohmic sections and the \marked{constriction region}.
	}
	\label{fig:voltage-characteristics-setup}
\end{figure}
In the previous subsection, we have considered an ideal de Laval nozzle with infinitely wide leads and without momentum relaxation (e.g,. Ohmic dissipation) between the contacts. In this subsection, we consider a more realistic setup, as shown in Fig.~\ref{fig:setup}. We consider a (2D graphene) de Laval nozzle \marked{geometry with a finite width up to the contacts and with Ohmic dissipation,} which is typically unavoidable (and can be expected to significantly exceed a chemical potential difference-induced $U_\mathrm{dL}$) in experimental transport setups~\cite{Geurs2025}. \marked{For this, we consider Ohmic sections with constant width to the left and right of the constriction region.}
By separating the Ohmics (only in the \marked{sections with constant width}) from the hydrodynamics (only in the \marked{constriction region}) in a lumped-element circuit model approach (see Figs.~\ref{fig:setup} and \ref{fig:voltage-characteristics-setup}), we apply our analytical solutions for the nozzle to this setup.

We consider the following width profile for the transport geometry (see Fig.~\ref{fig:voltage-characteristics-setup}a):
\begin{equation}
    W(x) = W_\mathrm{t} \frac{1 - r + \cosh(q L/2) + r \cosh(2 q x)}{\cosh(q L/2) + \cosh(2 q x)},
\end{equation}
with $L$ the length of the \marked{constriction geometry} (referred to below as \marked{$L_\mathrm{c}$} to distinguish it from the length of the Ohmic \marked{sections}), \marked{$r = W_\Omega/W_\mathrm{t}$} the ratio between the width of the Ohmic \marked{sections} [\marked{$W_\Omega \approx W(x)$} for \marked{$|x| \gtrapprox L_\mathrm{c}/2$}] and the width of the nozzle throat [$W_\mathrm{t} = W(x=0)$], and \marked{$q \gtrsim 1/L_\mathrm{c}$} controlling the steepness of the constriction profile of the nozzle (here always set to \marked{$q = 8/L_\mathrm{c}$}).
We consider geometrically identical Ohmic \marked{sections} to the left and right of the \marked{constriction} with length \marked{$L_\Omega$}, at chemical potentials $\mu_\mathrm{L}$ and $\mu_\mathrm{R}$, respectively, with the average $\langle \mu \rangle = (\mu_\mathrm{L} + \mu_\mathrm{R})/2$ kept fixed (through $\langle \mu \rangle = (\pi \hbar^2 v^2 n_\mathrm{2D})^{1/2}$, with $n_\mathrm{2D}$ the charge carrier density of the graphene sample).
The resistances of the \marked{Ohmic sections} $R_\Omega^{\Le,\Ri}$ are given by \marked{$R_\Omega^{\Le,\Ri} = L_\Omega \pi \hbar^2 v^2 / (W_\Omega \, e \, \mu_{\Le,\Ri}^2 \, \mu_\mathrm{mob})$}, with $\mu_\mathrm{mob}$ the mobility of the graphene sample.

We can thus consider the following relation that connects the overall voltage difference between the contacts to the chemical potential (or voltage) difference across the nozzle:
\begin{align} \label{eq:current-voltage-Ohmic}
    U_\mathrm{LR} &= (R_\Omega^{\Le} + R_\Omega^{\Ri}) (-e) I + (\mu_\mathrm{L} - \mu_\mathrm{R})/(-e) \\
    &= (R_\Omega^{\Le} + R_\Omega^{\Ri}) (-e) I + U_\mathrm{dL},
\end{align}
with $I$ the particle current through the transport geometry and $\mu_\mathrm{L}$ and $\mu_\mathrm{R}$ the chemical potentials in the left and right Ohmic \marked{sections}, which are matched to the chemical potentials at the left (\marked{$x=-L_\mathrm{c}/2$}) and right (\marked{$x=+L_\mathrm{c}/2$}) ends of the \marked{constriction} at finite cross section, respectively (see Fig.~\ref{fig:setup}).

\begin{figure*}[tb]
	\centering
	\includegraphics[width=.95\linewidth]{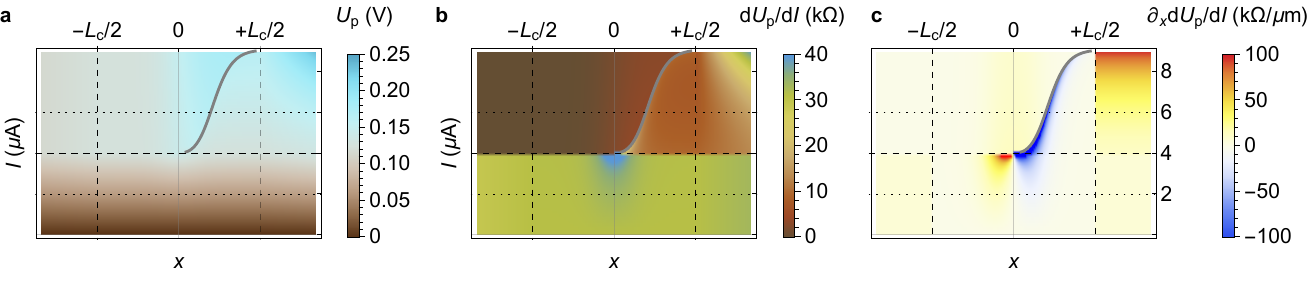}
	\caption{
		(\textbf{a})-(\textbf{c}) The normalized (\textbf{a}) probe voltage $U_\mathrm{p}$, (\textbf{b}) differential resistance $\mathrm{d} U_\mathrm{p}/\mathrm{d} I$, and (\textbf{c}) spatial derivative of the differential resistance $\partial_x \mathrm{d}U_\mathrm{p}/\mathrm{d} I$ as a function of the probe position $x$ and the current, for a nozzle with Ohmic \marked{sections with constant width} as shown in Figs.~\ref{fig:setup} and \ref{fig:voltage-characteristics-setup}.
        The horizontal dotted black lines indicate the currents of the subsonic and supersonic solutions shown in Fig.~\ref{fig:voltage-characteristics}.
        The gray line indicates the position of the shock front as a function of the current.
        The results are obtained for a transport geometry with \marked{$L_\Omega = 5\,\text{\textmu m}$, $W_\Omega = 1\,\text{\textmu m}$, $L_\mathrm{c} = 1\,\text{\textmu m}$}, $W_\mathrm{t} = 0.2\,\text{\textmu m}$, considering a graphene sample with $v =10^6\,\text{m/s}$, $n_\mathrm{2D} = 10^{11}\,\text{cm$^{-2}$}$ and $\mu_\mathrm{mob} = 10 000 \, \text{cm$^2$/(V$\cdot$s)}$.
        For this setup, we obtain $U_\mathrm{crit} = 0.25\,\text{V}$ ($e I_\mathrm{max} \approx 4\,\text{\textmu A}$) and the range of currents shown here corresponds to a range of $U_\mathrm{LR}$ from zero up to $1.25\,\text{V}$.
	}
	\label{fig:voltage-characteristics-ext}
\end{figure*}

As long as the current is below a critical current $I_\mathrm{crit}$, given by $I_\mathrm{max}$ in Eq.~\eqref{eq:voltage-characteristics-current}, the flow profile in the nozzle remains subsonic with $\mu_\mathrm{L} = \mu_\mathrm{R}$ and \marked{$R_\Omega^\Le = R_\Omega^\Ri \equiv R_\Omega = L_\Omega/(W_\Omega \, e \, n_\mathrm{2D} \mu_\mathrm{mob})$} such that the current-voltage relation is fully Ohmic: $U_\mathrm{LR} = 2 R_\Omega (-e) I$.
This behavior is maintained up to a critical bias voltage $U_\mathrm{crit} = 2 R_\Omega (-e) I_\mathrm{crit}$, at which the flow speed reaches the speed of sound at the nozzle throat.
When further increasing the voltage between the contacts above this critical value, the de Laval nozzle enters the supersonic regime with the voltage split over the Ohmic sections and the \marked{constriction region} according to Eq.~\eqref{eq:current-voltage-Ohmic} (see Fig.~\ref{fig:voltage-characteristics-setup}b).
In this regime, the voltage controls the chemical potential difference over the \marked{constriction} (with a certain voltage division) and thereby shifts the position of the shock front (see Fig.~\ref{fig:voltage-characteristics-ext}a).
%
Note that the lumped-element model breaks down above a certain bias voltage for which the width of the shock front position would exceed the width of the \marked{Ohmic sections} (typically still with $|U_\mathrm{dL}| \ll |U_\mathrm{LR}|$).

We can now also consider the local voltage characteristics within the nozzle geometry for this more realistic setup. For this, we introduce the voltage difference between the left contact and a specific position along the nozzle, here denoted as the probe voltage $U_\mathrm{p}(x)$:
\begin{equation}
    U_\mathrm{p}(x) = (R_\Omega^\Le + R_\Omega^\Ri) (-e) I + [\mu_\Le - \mu(x)]/(-e).
\end{equation}
Similar to the quantities introduced in the previous subsection, we can consider the local differential resistance, $\mathrm{d} U_\mathrm{p}(x)/\mathrm{d} I$, and its spatial derivative, $\partial_x \mathrm{d} U_\mathrm{p}(x)/\mathrm{d} I$. They are presented for a range of bias voltages across the subsonic-to-supersonic transition at the critical voltage $U_\mathrm{crit}$ in Fig.~\ref{fig:voltage-characteristics-ext}.

\section{Viscosity}
\label{sec:viscous_flow}
In this section, we consider the impact of (bulk) viscosity on the (supersonic) flow profiles in a Dirac electron nozzle. The viscosity can be described as an additional term in the Navier-Stokes equation, given by $-\zeta \Delta V$, with constant (bulk) viscosity $\zeta$~\cite{Narozhny2019}. Under the assumption of laminar flow, in which case the transverse component of the Laplacian of the viscosity term drops out ($\partial^2_\perp V = 0$) and the flow equation remains one-dimensional, viscous flow across the nozzle can be described with the following modified 1D Navier-Stokes equation [see Eq.~\eqref{eq:hydro_final}]:
\begin{equation}
	\partial P + M V \partial V = \zeta \partial^2 V,
\end{equation}
with viscosity $\zeta$.

With the viscosity term included, the 1D nozzle equation becomes:
\begin{equation} \label{eq:viscous_nozzle_eq}
	\frac{\partial V}{V} = -\frac{1 - (V/v)^2}{1 - (V/v_\sound)^2} \frac{\partial A}{A} - \xi A \frac{(V/v^2) \partial^2 V}{1 - (V/v_\sound)^2},
\end{equation}
writing the viscosity term in terms of an \emph{effective viscosity} $\xi \equiv (D + 1) \zeta / (M V A) = \text{const.} > 0$.
Note that considering this effective viscosity to be constant is equivalent to considering the \emph{physical viscosity} $\zeta$ to be constant due to the conservation of momentum [$\partial (M V A) = 0$]. This assumption thus boils down to neglecting any dependence of the viscosity on chemical potential and temperature (also see subsection on viscosity in graphene below), which vary across the nozzle in general.
Further note that $\xi$ is dimensionless for a two-dimensional ($D = 2$) nozzle, e.g., a graphene-based nozzle.

We numerically resolve Eq.~\eqref{eq:viscous_nozzle_eq} throughout the nozzle region, starting with boundary conditions for $V$ ($V = 0$) and $\partial V$ at the left lead. In practice, we propagate the solution from a very small but finite distance from the infinitely-wide left lead with different (finite) flow speeds $V$ and $\partial V = 0$. The acceleration of the flow readjusts to the proper boundary value over a very short distance and well-behaved subsonic and supersonic flow profiles are obtained, as can be seen in Fig.~\ref{fig:viscous_flow_profiles}.

\subsection{Viscous flow profile}
\label{subsec:viscous-flow-profile}
In the section above, we obtained the flow profile for a given initial flow speed at the left lead and a particular value of effective viscosity $\xi$. Here, we work out how the viscous flow profile changes as a function of the pressure difference across the nozzle. As in the perfect-fluid regime, this pressure difference can be obtained by applying a voltage or temperature difference between the (infinitely-wide) leads of the nozzle, where the viscosity is effectively zero. We can relate the pressure difference to the flow profile at the nozzle ends by considering momentum conservation and evaluating it at the nozzle ends where the flow speed vanishes as follows:
\begin{equation}
\begin{split}
	\left.(M V A)\right|_\Le &= \left.(M V A)\right|_\Ri \\
    \Rightarrow \frac{P_\Ri}{P_\Le} &= \left.\left| \frac{\partial V}{\partial (1/A)} \right|_\Le \right/ \left| \frac{\partial V}{\partial (1/A)} \right|_\Ri,
\end{split}
\end{equation}
making use of the expression for pressure in Eq.~\eqref{eq:macrovalues_eq} and of the relation $|V A| = |\partial V / \partial(1/A)|$ in the limit $V \rightarrow 0$, $A \rightarrow + \infty$.
We can also relate the pressure inside the nozzle to the pressure in the left lead, making use of the same expression for the pressure:
\begin{equation}
	\frac{P}{P_\Le} = \left(1 - \frac{V^2}{\upsilon^2} \right) \frac{\left| \partial V/\partial (1/A) \right|_\Le}{V A}
\end{equation}
Now, let us consider the chemical potential and temperature (and thus pressure) of the left lead fixed, as well as the physical viscosity, while increasing the flow speed by letting the pressure drop at the right lead.
As the viscosity has a proportionality $\zeta \propto \xi \partial V_\Le$, we must simultaneously rescale $\xi$ with a factor $\partial V_\Le$ while lowering the pressure on the right to keep the physical viscosity $\zeta$ constant. This implies that the effective viscosity in the viscous nozzle equation decreases as the flow speed increases (by lowering pressure on the right lead) and vice versa.

\begin{figure*}[tb]
	\centering
	\includegraphics[width=\linewidth]{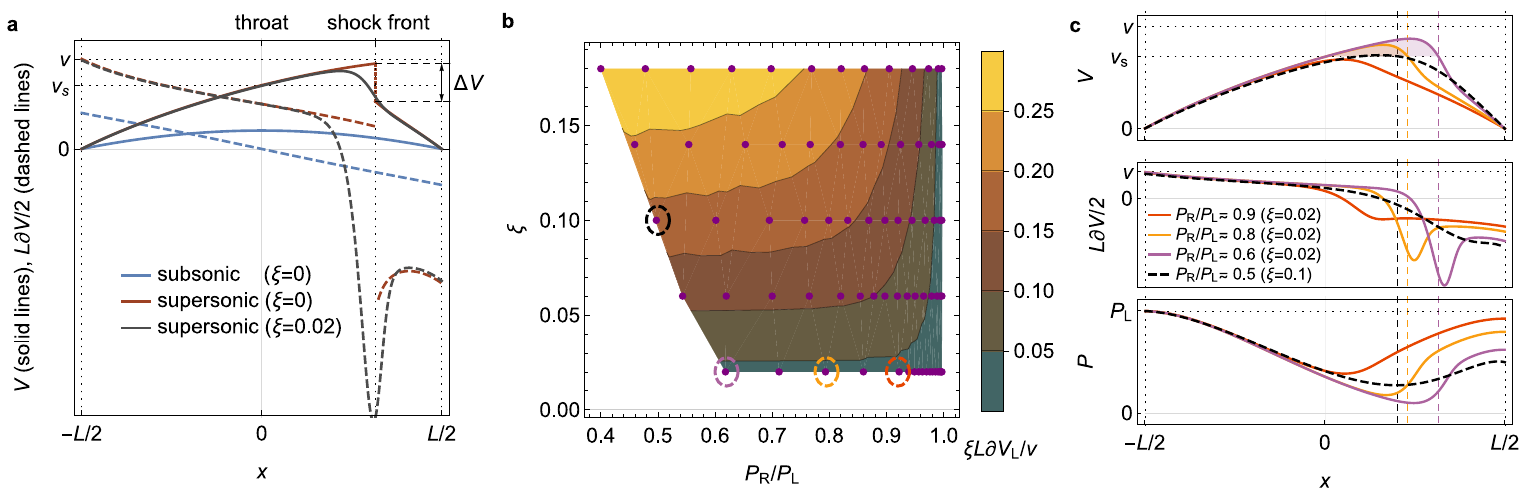}
	\caption{
		(\textbf{a}) The flow speed $V$ (solid lines) and acceleration $\partial V$ (dashed lines) of a supersonic viscous flow profile of the two-dimensional nozzle profile considered in Fig.~\ref{fig:distribution}, considering effective viscosity $\xi = 0.02$. A subsonic and a supersonic flow profile, considering $\xi = 0$ (corresponding to the perfect-fluid regime), are shown for comparison.
		(\textbf{b}) The product of effective viscosity $\xi$ and acceleration at the left lead $\partial V_\Le$ is shown as a function of the pressure ratio across the nozzle $P_\Ri/P_\Le$ and the effective viscosity $\xi$. The isocontours of this product correspond to solutions with constant viscosity $\zeta$. The data points from which the color map was obtained through interpolation are indicated by purple dots.
		(\textbf{c}) The flow speed, acceleration, and pressure profiles of four different (near-)supersonic flow profiles indicated in \textbf{b} by a dashed circle with the same color as the corresponding profiles, considering the same two-dimensional nozzle profile as in \textbf{a} and Fig.~\ref{fig:distribution}: $A(x) = A_\throat/[1-(2x/L)^2]$.
	}
	\label{fig:viscous_flow_profiles}
\end{figure*}

By mapping out the flow profiles in the two-dimensional parameter space of $\partial V_\Le$ (or, alternatively, $P_\Ri/P_\Le$) and viscosity parameter $\xi$, we can trace the flow-profile solutions along isocontours for constant physical viscosity $\zeta$ ($\propto \xi \partial V_\Le$) with increasing pressure difference across the nozzle. Such isocontours are presented in Fig.~\ref{fig:viscous_flow_profiles}b. It can be seen that $\xi$ becomes constant at higher pressure differences (flow speeds) along the isocontours. For low pressure differences, however, the effective viscosity shoots up.
In summary, the flow profile becomes effectively less viscous for higher flow speeds and drops down to a minimum effective viscosity that is determined by the specifics of the nozzle, the properties of the left lead (in particular, pressure of the Dirac electrons, or temperature and chemical potential), and the viscosity $\zeta$.

The flow speed, acceleration, and pressure profiles of different (near-)supersonic viscous flow profiles are presented in Fig.~\ref{fig:viscous_flow_profiles}c. While there are no more discontinuities in the viscous supersonic flow profiles, a clear remnant of the discontinuous shock front can still be seen when the effective viscosity parameter is small enough ($\xi \approx 0.02$), with a steep drop of the flow speed (supersonic to subsonic) and a steep upturn of the pressure. The remnant can be seen most clearly in the flow acceleration profile with a highly peaked deceleration where the flow profile returns from supersonic to subsonic flow speeds past the throat. Overall, the flow profiles in the (effectively) low-viscosity regime are in good qualitative and quantitative agreement with those obtained in the perfect-fluid regime.

\subsection{Viscosity in graphene}
\label{subsec:viscosity_graphene}
Evaluating the expression for the viscosity parameter at the infinitely-wide leads of the nozzle, we obtain:
\begin{equation} \label{eq:viscosity_dimensionless}
	\xi = \frac{(D + 1) \zeta}{M_0 } \left| \frac{\partial(1/A)}{\partial V} \right| \sim \frac{3 \zeta}{M_0 v A_\throat}.
\end{equation}
The viscosity $\zeta$ can be related to the microscopic interparticle scattering time $\tau_{e\text{-}e}$ as follows~\cite{Lucas2018}:
\begin{equation} \label{eq:viscosity_microscopic}
	\zeta \sim M_0 v^2 \tau_{e\text{-}e}.
\end{equation}
For graphene in the Dirac-fluid regime ($\mu \ll T$) at $T = 100\,\textnormal{K}$, we obtain~\cite{Lucas2018}:
\begin{align}
	\tau_{e\text{-}e} &\sim \frac{\hbar}{\kB T} \approx 7. \times 10^{-14}\,\textnormal{s}, \\
	M_0 &\approx \frac{3}{2 \pi} \frac{1.2 \times 3 (\kB T)^3/2}{v^4} \approx 2.0 \times 10^{-19}\,\textnormal{kg/m}^2, \\
	\zeta &\sim 1.6 \times 10^{-20}\,\textnormal{kg/s}.
\end{align}

Considering Eqs.~\eqref{eq:viscosity_dimensionless} and \eqref{eq:viscosity_microscopic}, the viscosity parameter becomes:
\begin{equation}
	\xi = 3 v \tau_{e\text{-}e} / A_\throat.
\end{equation}
For $\xi \ll 1$, we require that the width of the nozzle throat is much larger than the electron-electron scattering length $l_{e\text{-}e} = v \tau_{e\text{-}e} \sim 70\,\textnormal{nm}$.
Note that we also require the length of the nozzle to be smaller than the momentum relaxation length $l_\text{mom} = v \tau_\text{mom}$.
Considering the current experimental status in graphene, with $l_\text{mom}$ reaching values up to $1\,\textnormal{\textmu m}$, there should be a window of opportunity with nozzle dimensions between $100\,\textnormal{nm}$ and $1\,\textnormal{\textmu m}$ for resolving a signature of supersonic hydrodynamic flow with the appearance of a shock front.

\begin{widetext}
\section{List of symbols}
\label{sec:notation}

\begin{center}

  \begin{tabularx}{\linewidth}{|l X| l X|}
		\hline
		$\Ham$ & Dirac Hamiltonian &
		$D$ & number of spatial dimensions \\
		\hline
		$v$ & Dirac velocity &
		$v_\sound$ & speed of sound\\
		\hline
		$\vecp$ & momentum &
		$\vecsigma$ & vector of Pauli matrices \\
		\hline 
		$\lambda (\pm 1)$ & chirality or nature of particle (electron or hole) with intensive thermodynamic conjugate variable $\chi \equiv - X / T$ &
		$E$ & macroscopic energy density \\
		\hline
		$l_{e\textnormal{-}e}$ & interparticle scattering length &
		$l_\mathrm{mom}$ & typical length scale of momentum relaxation \\
		\hline
		$M$ & effective fluid mass density, with subscript 0 for vanishing flow velocity &
		$N$ & particle number (density), with subscript 0 for vanishing flow velocity \\
		\hline
		$I$ & particle current &
		$e$ & elementary charge \\
		\hline
		$P$ & pressure &
		$P_*$ & critical pressure (transition from subsonic to\newline supersonic flow) \\
		\hline
		$\Delta P$ & pressure jump at shock front position &
		$\xi$ & ratio of cross section of throat and cross section at certain position along nozzle \\
		\hline
    \end{tabularx}
    \begin{tabularx}{\linewidth}{|l X| l X|}
        \hline
		$\Xi$ & ratio of pressure drop versus pressure in left lead (where flow originates) &
		$\vecV$ ($V$) & flow velocity (speed) \\
        \hline
		$A$ & cross section of nozzle profile &
        $W$ & width of 2D nozzle profile \\
        \hline
		$\mu$ & chemical potential &
		$\Delta \mu$ & chemical potential difference (between opposite ends of nozzle or shock front) \\
        \hline
		$T$ & temperature &
		$C_{\mu, T, P}$ & solution constants of the nozzle equations\newline representing chemical potential, temperature, and pressure for vanishing flow speed \\
		\hline
		$C_A$ & solution constant of the nozzle equation that \newline relates cross section to flow speed &
		$\kappa$ & upper bound for $C_A / A$ \\
		\hline
		$\vecS (S)$ & macroscopic momentum (energy flow) &
		$f_\lambda(\vecp)$ & electronic distribution function \\
		\hline
		$f_{\Fermi/\Hydro}$ & Fermi-Dirac/hydrodynamic flow distribution function &
		$\Li_n$ & polylogarithm functions \\
		\hline
		$\vecn$ & propagation direction of electrons &
		$\rttensor{\Pi}$ & stress tensor \\
		\hline
		$\Lambda$ & macroscopic chirality  &
		$\vecj_\Lambda$ & chiral current \\
		\hline
		$S^D$ & surface of a $D$-dimensional sphere &
        $L$ & length of nozzle \\
		\hline
		$\tau_{e\textnormal{-}e}$ & typical interparticle collision time & $\tau_\text{mom}$ & typical time scale for momentum relaxation \\
		\hline
        $\zeta$ & viscosity & $\xi$ & (dimensionless) viscosity parameter \\
		\hline
		$\vecE$ & electric field & $\vecB$ & magnetic field \\
		\hline
		$U$ & voltage &
		$R$ & resistance \\
		\hline
    \end{tabularx}
\end{center}
\end{widetext}

\bibliography{references}

\end{document}